\documentclass{emulateapj}
\usepackage{float}
\usepackage{amsmath}

\begin{document}

\title{The Stellar Mass Fundamental Plane and Compact Quiescent Galaxies at $\lowercase{z}<0.6$}

\author{H. Jabran Zahid$^{1}$\footnote{email: zahid@cfa.harvard.edu}, Ivana Damjanov$^{1}$, Margaret J. Geller$^{1}$, Ho Seong Hwang$^{2}$ \& Daniel G. Fabricant$^{1}$}
\affil{$^{1}$Harvard-Smithsonian Center for Astrophysics - 60 Garden Street, Cambridge, MA 02138}
\affil{$^{2}$School of Physics, Korea Institute for Advanced Study, 85 Hoegiro, Dongdaemun-gu, Seoul 130-722, Republic of Korea}

\def\mean#1{\left< #1 \right>}

%\date{}                                           % Activate to display a given date or no date
%\begin{document}
\begin{abstract}
We examine the evolution of the relation between stellar mass surface density, velocity dispersion and half-light radius$-$the stellar mass fundamental plane$-$for quiescent galaxies at $z<0.6$. We measure the local relation from galaxies in the Sloan Digital Sky Survey and the intermediate redshift relation from $\sim500$ quiescent galaxies with stellar masses $10 \lesssim \mathrm{log}(M_\ast/M_\odot) \lesssim 11.5$. Nearly half of the quiescent galaxies in our intermediate redshift sample are compact. After accounting for important selection and systematic effects, the velocity dispersion distribution of galaxies at intermediate redshifts is similar to galaxies in the local universe. Galaxies at $z<0.6$ appear to be smaller ($\lesssim0.1$ dex) than galaxies in the local sample. The orientation of the stellar mass fundamental plane is independent of redshift for massive quiescent galaxies at $z<0.6$ and the zero-point evolves by $\sim 0.04$ dex. Compact quiescent galaxies fall on the same relation as the extended objects. We confirm that compact quiescent galaxies are the tail of the size and mass distribution of the normal quiescent galaxy population. 

%The similar size distributions of local and intermediate redshift quiescent galaxies limit size growth of the compact population to $\lesssim0.1$ dex or require that any compact galaxies that grow substantially are replaced by new compact galaxies. 

\end{abstract}

\section{Introduction}

In the $\Lambda$CDM paradigm, galaxies form as gas cools and condenses on to dark matter halos. As far as we know, the baryons which compose the observable parts of galaxies are influenced by dark matter only through gravitational interactions. Thus, observations of the structural and dynamical properties of galaxies are important for constraining the physical processes governing galaxy formation and evolution.  

Stellar velocity dispersion is an observable property directly connecting the baryonic content of galaxies to the unobservable dark matter halos \citep{Wake2012, Schechter2015}. Using the virial theorem, the velocity dispersion can be combined with galaxy size to yield an estimate of the dynamical mass \citep[e.g.][]{Cappellari2006}; the dynamical mass is proportional to the total mass of the system \citep{Bolton2008}. The dynamical mass is also correlated with the baryonic content of galaxies which is proportional to the luminosity and/or stellar mass. 

 %Observations of the structural and dynamical properties of galaxies provide a unique probe for studying galaxies and their underlying dark matter halos and thus provide important constraints for models of galaxy formation and evolution.

Quiescent galaxies in the local universe exhibit a tight relation between size, velocity dispersion and surface brightness/stellar mass surface density \citep{Djorgovski1987, Dressler1987, Bernardi2003, Hyde2009b, Saulder2013}. This relation extends across a broad range of early-type galaxies in the local universe \citep{Misgeld2011}. The relation, described by the so-called fundamental plane (FP), reflects the virial equilibrium of galaxies. The FP evolves with redshift \citep{Treu2005, vanDokkum2007,Holden2010, Saglia2010, vandeSande2014, Zahid2015a}; this evolution is attributed mainly to evolution in the mass-to-light (M/L) ratio of galaxies. Replacing the surface brightness with stellar mass surface density yields a stellar mass fundamental plane (MFP). Because stellar mass estimates ostensibly account for M/L ratio variations, changes in the MFP should trace evolution of the structural or dynamical properties of galaxies. Very little evolution in the MFP is observed out to $z\sim2$ \citep{Bezanson2013}. This conclusion is based on a small number of observations because of the difficulty in obtaining the high signal-to-noise spectroscopy required for measuring velocity dispersion.

The FP is a tool for studying the evolution of galaxies. \citet[Z15 hereafter]{Zahid2015a} examine the FP of quiescent galaxies with imaging from the \emph{Hubble Space Telescope (HST)} and spectroscopy from the Sloan Digital Sky Survey (SDSS) out to $z\sim0.8$. The SDSS sample is strongly biased towards the highest surface brightness objects and thus a majority of the galaxies in the sample are compact, i.e. they are several times smaller than typical quiescent galaxies at a fixed stellar mass. These galaxies are offset from the local FP but passive evolution alone is enough to bring them onto the local relation by $z\sim0$. Thus, the FP is a constraint on the evolution of the structural and dynamical properties of compact galaxies; these objects remain on the relation as they evolve.

Studies based on the SDSS suggest that the number density of compact quiescent galaxies declines significantly at $z<1$ \citep{Trujillo2009, Taylor2010} and the average size of quiescent galaxies increases as the universe evolves \citep{Daddi2005, Toft2007, Trujillo2007, Zirm2007, Buitrago2008, vanDokkum2008b,Damjanov2011, Newman2012, Cassata2013, vanderWel2014}. Based on these observations, several growth scenarios have been proposed for the compact galaxy population \citep[e.g.,][]{Khochfar2006, Fan2008, Naab2009}. However, in contrast to early estimates based on SDSS, studies of high density regions in the local universe \citep{Valentinuzzi2010, Poggianti2013} and recent measurements at intermediate redshift \citep{Carollo2013, Damjanov2014, Damjanov2015a} show that the number density of compact quiescent galaxies does not evolve significantly at $z<1$. To maintain an approximately constant number density of compact galaxies at $z<1$ as recent studies suggest either 1) compact galaxies do not grow at $z<1$ and the larger average size measured for quiescent galaxies in the local universe is solely a result of the addition of new, large galaxies into the quiescent galaxy population, i.e. progenitor bias \citep[e.g.,][]{Carollo2013}, or 2) that growth in the quiescent compact galaxy population is balanced by new compact galaxies forming at $z<1$.

Here we extend the work of Z15 using new measurements based on observations obtained using Hectospec on the \emph{MMT}. We measure the structural and dynamical properties of galaxies in the COSMOS field and examine the evolution of the MFP of quiescent galaxies at $z<0.6$. A large fraction of these galaxies are compact. In Section 2 we describe the data and methods. In Section 3 we address selection and systematic effects and Section 4 contains our derivation of the stellar mass fundamental plane. We discuss our results in Section 5 and conclude in Section 6. Throughout we assume the standard cosmology $(H_{0}, \Omega_{m}, \Omega_{\Lambda})$ = (70 km s$^{-1}$ Mpc$^{-1}$, 0.3, 0.7), AB magnitudes and a \citet{Chabrier2003} IMF.

\section{Data and Methods}

\subsection{Observations}

We derive the MFP from a sample of local galaxies taken from the Sloan Digital Sky Survey \cite[SDSS;][]{Alam2015}. We restrict our sample to the SDSS Legacy Survey which includes spectroscopy of $\sim900,000$ galaxies with $r<17.8$ \citep{York2000}. The nominal spectral range of the observations is $3800 - 9200 \mathrm{\AA}$ and the spectral resolution is $R\sim1500$ at 5000$\mathrm{\AA}$ \citep{Smee2013}. Five bands of optical photometry ($ugriz$) are available for the whole spectroscopic sample \citep{Doi2010}. Throughout this work, we use the c-model AB magnitudes.

We examine the redshift evolution of the MFP using data from the 1.6 deg$^2$ COSMOS\footnote{http://cosmos.astro.caltech.edu/} field \citep{Scoville2007}. We have conducted a redshift survey of the field \cite[Damjanov et al., In Prep]{Damjanov2015b} using Hectospec \citep{Fabricant2005} on the \emph{MMT} to observe $\sim2500$ objects selected from the UltraVISTA catalog\footnote{http://www.strw.leidenuniv.nl/galaxyevolution/ULTRAVISTA/ Ultravista/K-selected.html} of \citet{Muzzin2013a}. The survey is designed to observe early-type galaxies efficiently. Our targets are color selected ($g - r > 0.8$, $r-i > 0.2$) in the magnitude range $17.8<r<21.3$. The color selection is sufficiently broad to completely target the quiescent galaxy population \citep[see][]{Damjanov2015b}. The bright limit of our survey is the SDSS magnitude limit. The nominal spectral range of our observations is $3700-9150 \mathrm{\AA}$ and the spectral resolution is $R\sim1000$ at 5000$\mathrm{\AA}$. We adopt the $ugriz-$band photometry given in the \cite{Muzzin2013a} catalog. We refer to these data as the hCOSMOS sample\footnote{A small subset of the galaxies in the hCOSMOS sample are from SDSS because we chose not to observe galaxies in the COSMOS field brighter than the SDSS magnitude limit.}. 

Our typical integration time was $\sim1$h and the data were obtained under variable conditions in February and April of 2015. We measured 2096 new redshifts in the COSMOS field at  $0.1 \lesssim z \lesssim0.6$ (Damjanov et al., In Prep). We can extract velocity dispersion for only a subset of our observations; this measurement requires higher signal-to-noise (S/N) spectroscopy than measurements of redshift. We measure velocity dispersion at a S/N $>3$ and a reduced $\chi^2 < 2$ for $\sim800$ galaxies in the sample. The typical S/N per resolution element is $\sim7$ at $5000 \mathrm{\AA}$ for the subset of the sample for which we are able to reliably measure velocity dispersion. 

We identify quiescent galaxies based on broad-band colors and the $D_n4000$ index (see below). \citet{Balogh1999} define $D_n4000$ as the ratio of the flux in two windows adjacent to the 4000$\mathrm{\AA}$ break. The $D_n4000$ index is sensitive to age of the stellar population and shows a clear bimodal distribution which separates the star-forming and quiescent galaxy population out to intermediate redshifts \citep{Kauffmann2003a, Geller2014}.

Various definitions for compactness can be found in the literature. We use the \cite{Barro2013} classification for compactness 
\begin{equation}
 \frac{M_\ast}{R^{1.5}_e} \geq 10^{10.3} \, \left[ \mathrm{M_\odot} \, \mathrm{kpc}^{-1.5} \right].
\end{equation}
Here, $R_e$ is the circularized half-light radius measured in kpc and $M_\ast$ is the stellar mass measured in solar mass units. We only classify galaxies with \emph{HST} imaging (hCOSMOS sample) using this compactness criteria. Based on this classification, 45\% (228/509) of the galaxies in our selected sample of quiescent objects with velocity dispersions are compact.

\begin{figure*}
\begin{center}
\includegraphics[width=1.8\columnwidth]{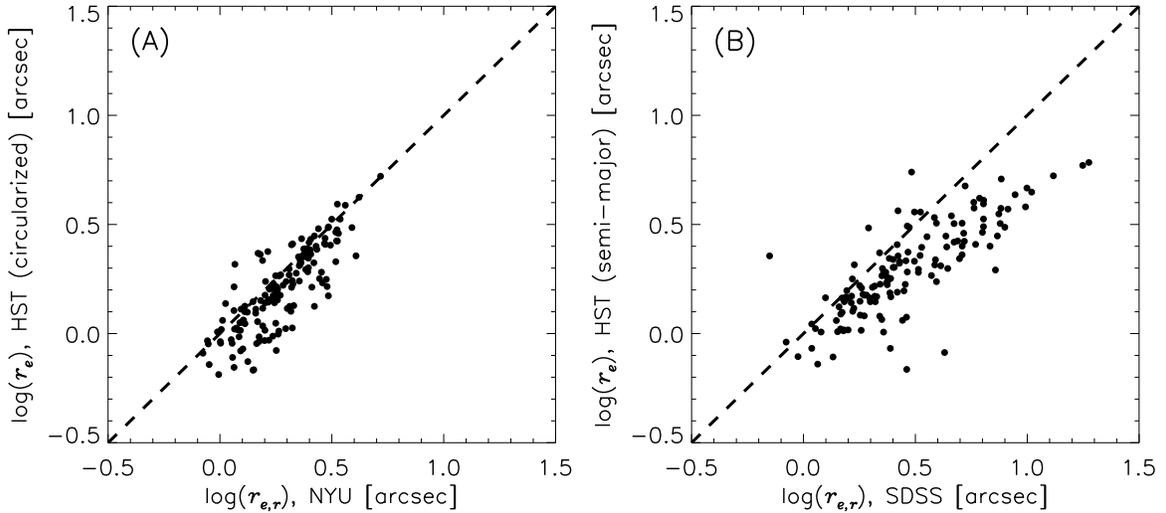}
\end{center}
\caption{Comparison between angular sizes measured using \emph{HST} ACS imaging and SDSS photometry. The SDSS $r-$band sizes are (A) measured by the NYU group and (B) an output of the SDSS photometric pipeline. The de Vaucouleurs profile fits from the SDSS photometric pipeline are reported as the semi-major axis, thus we do not circularize the HST radii in (B). The dashed line shows the one-to-one agreement.}
\label{fig:size}
\end{figure*}

\subsection{Methods}

\subsubsection{Stellar Masses}

We measure stellar masses for the SDSS and hCOSMOS sample using the LePHARE\footnote{http://www.cfht.hawaii.edu/$\sim$arnouts/LEPHARE/lephare.html} spectral energy distribution (SED) fitting code \citep{Arnouts1999, Ilbert2006b}. LePHARE calculates the mass-to-light ratio by fitting synthetic SEDs to observed photometry. We use the stellar population synthesis models of \citet{Bruzual2003}. The models have two metallicities. Synthetic SEDs are generated by varying the star formation history, age, and extinction of the stellar population. The star formation histories are exponentially declining ($\propto e^{-t/\tau}$) with $e$-folding times ($\tau$) ranging between 0.1 and 30 Gyr. We adopt the \citet{Calzetti2000} extinction law with $E(B-V)$ ranging from $0 - 0.6$ and the \citet{Chabrier2003} initial mass function. We take the median of the stellar mass distribution as our estimate of the stellar mass. We also generate absolute, $k-$corrected $ugriz$ magnitudes by convolving the filter functions with the best-fit SED.

\subsubsection{Sizes}

Sizes of hCOSMOS galaxies are measured by \citet[S07 hereafter]{Sargent2007} from \emph{Hubble Space Telescope (HST)} imaging of the COSMOS field \citep{Koekemoer2007}. The images are taken with the Advanced Camera for Surveys (ACS) using F814W filter. S07 fit the two-dimensional surface brightness profile with a single \citet{Sersic1968} profile using GIM2d \citep{Simard2002}. The half-light radius and Sersic index are free parameters of the model. The \emph{HST} observations have a resolution $<0''\!\!.1$ \citep{Koekemoer2007}. The mean angular half-light radius, $r_e$, of galaxies in the hCOSMOS sample is $0''\!\!.8$ and 99\% of galaxies in the sample have half-light radii $\gtrsim0''\!\!.25$.

The minimum measurable size for SDSS galaxies is limited by the typical seeing of $>1''\!\!.2$ \citep{Stoughton2002}. This corresponds to the minimum diameter, limiting the minimum effective radius to half this value. SDSS sizes have been measured using several different procedures. Thus, the hCOSMOS sizes based on \emph{HST} imaging may differ systematically from SDSS sizes. The COSMOS field overlaps the SDSS footprint. Half-light radii measured by S07 from \emph{HST} imaging are available for $\sim150$ galaxies in the SDSS photometric catalog. In Figure \ref{fig:size} we compare two independent measurements of half-light radius based on SDSS imaging with measurements from S07. Figure \ref{fig:size}A shows size measurements taken from the NYU value added catalog \citep{Blanton2005a, Padmanabhan2008}. These measurements\footnote{These are designated as SERSIC\_R50 in the catalog.} are based on Sersic profile fits where the index is a free parameter; similar to the approach used by S07 for measuring sizes based on \emph{HST} imaging. The NYU size estimates are systematically larger by 0.071 dex but show no significant trend with size for the range of sizes compared in Figure 1A. The systematic offset may be due to the poorer seeing of the SDSS ground based photometric data and/or to the methodology used to measure sizes. Figure \ref{fig:size}B shows measurements based on de Vaucouleurs profile fits\footnote{These are designated THETA\_DEV in the SDSS photoObj output.} \citep{Stoughton2002} used in several studies of the FP of SDSS galaxies \citep[e.g.,][]{Bernardi2003, Hyde2009b}. The half-light radii measured from \emph{HST} imaging by S07 are clearly more consistent with the half-light radii measured by the NYU group (Figure \ref{fig:size}A) as is expected since the procedure for these two measurements is very similar. Thus, we adopt the NYU $r-$band measurements as the half-light radii for the SDSS sample but subtract 0.071 dex to account for the systematic offset. Based on our comparison we adopt a fiducial uncertainty of 0.1 dex for the NYU size measurements. This correction does not significantly effect our results.

Galaxy sizes depend on the observed wavelength. \citet{vanderWel2014} find that for early-type galaxies, the change in size as a function of wavelength is given by $\Delta \mathrm{log}(R) / \Delta \mathrm{log} \lambda$ = -0.25 where $R$ is radius and $\lambda$ is wavelength. Because we examine a large redshift range, we correct all galaxy sizes to rest-frame 6030$\mathrm{\AA}$ using the relation derived by \citet{vanderWel2014}. This rest-frame wavelength corresponds to the effective rest-frame wavelength of the \emph{HST} ACS F814W filter at the median redshift of the hCOSMOS sample (z = 0.35). 

Measurements of the half-light radii by S07 are reported as the semi-major half-light radius. We convert this to a circularized half-light radius by multiplying the semi-major half light radius by $\sqrt{b/a}$ where $a$ and $b$ are the semi-major and semi-minor axis, respectively. The NYU sizes are inherently circularized by their measurement procedure \citep[see appendix of ][]{Blanton2005b}. 

The median correction for wavelength dependent size is $-0.004$ dex for the SDSS sample. The median correction for wavelength dependent size and circularization is -0.08 dex for the hCOSMOS sample. We convert the circularized half-light radius measured in arc seconds to a physical half-light radius, $R_e$, measured in kpc using the relation between angular diameter distance and redshift.

We calculate the stellar mass surface density
\begin{equation}
\Sigma_\ast = \frac{M_\ast}{2 \pi R_e^2}
\end{equation}
in units of $M_\odot$ kpc$^{-2}$. Here, $M_\ast$ is the stellar mass and we assume that $R_e$ contains half of of the stellar mass. This assumption may not strictly be true due to potential M/L ratio gradients.

\subsubsection{Velocity Dispersions}

The stellar velocity dispersion of SDSS galaxies are measured by \citet{Thomas2013} using the Penalized PiXel Fitting \citep[pPXF][]{Cappellari2004} and Gas and Absorption Line Fitting \citep[GANDALF][]{Sarzi2006} codes. The stellar population and emission line templates are simultaneously fit to the observed spectrum. The stellar line-of-sight velocity dispersion is fit in pixel space using pPXF and the \citet{Maraston2011} stellar population models matched to the SDSS resolution. The spectral resolution, redshift of targets and the S/N of the observations sets the minimum velocity dispersion that can be reliably measured. For SDSS this typically corresponds to $\sim60$ km s$^{-1}$ \citep{Thomas2013}

We measure the hCOSMOS velocity dispersions using the University of Lyon Spectroscopic analysis Software\footnote{http://ulyss.univ-lyon1.fr/} \citep[ULySS;][]{Koleva2009}. Single age stellar population models calculated with the PEGASE-HR code \citep{LeBorgne2004} and the MILES stellar library \citep{Sanchez-Blazquez2006} are used to fit the observed spectrum. Our sample selection criterion of $D_{n}4000<1.5$ (see Section 3.1) effectively removes emission line galaxies from the sample \citep{Woods2010}. The spectral fits are limited to $4100-5500$ $\mathrm{\AA}$ rest-frame wavelengths. This spectral range yields the smallest velocity dispersion errors and the most stable results \citep{Fabricant2013}. This spectral range is accessible with Hectospec throughout the redshift range spanned by the hCOSMOS data. Based on the measured line spread function, models are convolved to the wavelength dependent spectral resolution of the Hectospec data \citep{Fabricant2013}. The models are parameterized by age and metallicity and are convolved with the line-of-sight velocity dispersion. The best-fit age, metallicity and velocity dispersion are determined by $\chi^{2}$ minimization. Given the redshift and S/N of our observations and the slightly lower resolution of Hectospec, we empirically determine that velocity dispersions are typically reliable down to $\sim90$ km s$^{-1}$. Details of the velocity dispersions measurements with Hectospec and systematic issues related to the measurement are discussed in \citet{Fabricant2013}.

The velocity dispersion varies with aperture size. \citet{Cappellari2006} find  
\begin{equation}
\frac{\sigma_1}{\sigma_2} = \left(\frac{r_1}{r_2}\right)^{-0.066}, 
\label{eq:cap}
\end{equation}
where $\sigma_1$ and $\sigma_2$ are velocity dispersions measured in apertures of radii $r_1$ and $r_2$, respectively. The fiber apertures for Hectospec and SDSS are $0''\!\!.75$ and $1''\!\!.5$, respectively. We correct the measured velocity dispersion to the effective angular radius, $r_e$, using Equation \ref{eq:cap}. The median correction is 0.003 and 0.005 dex for the SDSS and hCOSMOS samples, respectively.

\begin{figure}
\begin{center}
\includegraphics[width=\columnwidth]{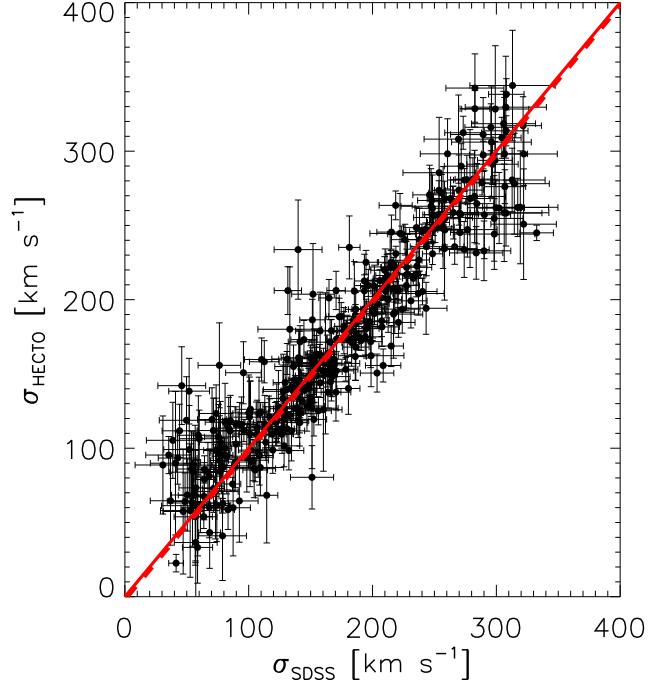}
\end{center}
\caption{Comparison of velocity dispersions measured using Hectospec on \emph{MMT} and the SDSS spectrograph. The solid line is the one-to-one correspondence and the dashed line is the best fit relation.}
\label{fig:vdisp}
\end{figure}

To assess systematic uncertainties in the determination of velocity dispersions, we compare measurements made with Hectospec to SDSS measurements \citep{Thomas2013}. The SHELS F2 survey is a highly complete redshift survey of four deg$^{2}$ conducted with Hectospec \citep{Geller2005, Geller2014}. The survey overlaps with the SDSS footprint and several hundred galaxies are observed in both surveys \citep[see also][]{Fabricant2013}. We identify galaxies in the SHELS F2 field that are also observed by the SDSS. Velocity dispersions for the SHELS F2 and SDSS galaxies are measured using the procedures for Hectospec data and SDSS outlined above, respectively. We directly compare the two measurements to assess systematic differences in the determination of velocity dispersions. 

Figure \ref{fig:vdisp} shows the comparison of the velocity dispersion measured for the same objects by the two different instruments and procedures. Prior to making the comparison, a small correction is applied to account for the differing fiber apertures \citep[see Equation \ref{eq:cap}]{Fabricant2013}. We fit a line to the relation between the two measurements using the \emph{fitexy.pro} routine in IDL. This fit accounts for errors in both coordinates. The best fit relation is
\begin{equation}
\sigma_{\mathrm{HECTO}} = (-2.69 \pm 0.83) + (1.00 \pm 0.01) \sigma_{\mathrm{SDSS}} \, \mathrm{[km \, s^{-1}]}.
\end{equation}
The fit demonstrates that the two measurements are relatively consistent and systematically offset by 2.7 km s$^{-1}$. We subtract 2.7 km s$^{-1}$ from the SDSS velocity dispersion measurements. The root-mean-square (RMS) dispersion of the two measurements is 26 km s$^{-1}$ and the intrinsic uncertainty (accounting for observational uncertainty in both measurements) is 16 km s$^{-1}$ \citep[c.f.,][]{Fabricant2013}. The 2.7 km s$^{-1}$ correction is significantly smaller than the intrinsic uncertainty; this small correction to velocity dispersion does not affect any of our results.

\section{Data Selection and Systematics}

Our aims are to quantify evolution in the structural and kinematic properties of quiescent galaxies using the MFP and to compare these properties for the compact and extended quiescent galaxy population. 

In this section we address two key issues prior to deriving the MFP: 
\begin{enumerate}

\item Structural and kinematic properties of galaxies are correlated with various observational selection effects. Thus, selection effects may introduce bias and spurious trends with redshift

\item Differences in the distributions of velocity dispersions and sizes may result from differences in the error distributions and systematic limitations in the sensitivity of the measurements. The hCOSMOS spectroscopy can only reliably probe down to $\sim90$ km s$^{-1}$ whereas the SDSS observations are typically reliable down to $\sim60$ km s$^{-1}$. The SDSS photometry is seeing limited typically at $\sim1''\!\!.2$ whereas \emph{HST} imaging is diffraction limited at $<0''\!\!.1$. Thus, these two systematic limitations artificially truncate the tails of the velocity dispersion and size distributions of the hCOSMOS and SDSS samples, respectively. 
\end{enumerate}
Here we consider these effects and take them into account.

\begin{figure*}
\begin{center}
\includegraphics[width=2\columnwidth]{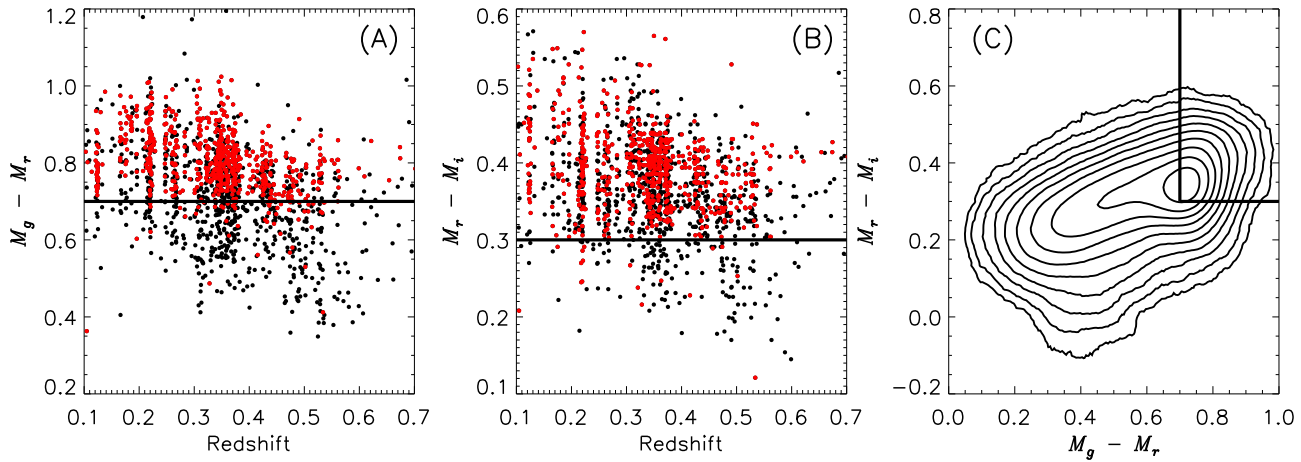}
\end{center}
\caption{Rest-frame (A) $g-r$ and (B) $r-i$ colors of the hCOSMOS galaxies as function of redshift. The red data points are galaxies with $D_n4000 > 1.5$. The solid lines are the empirically determined color cuts which primarily select the quiescent galaxy population. The cuts are based on examination of the $D_n4000$ distribution. (C) Contours of the $g-r$ vs. $r-i$ color distribution of SDSS galaxies. The solid lines indicate the color cuts in (A) and (B).}
\label{fig:color}
\end{figure*}

\subsection{Data Selection}

The redshift range of the SDSS data we use is restricted to $z<0.1$. This redshift limit makes the sample large enough that statistical errors do not obscure evolutionary trends. Furthermore, sizes can be accurately estimated for a large fraction of the sample; sizes become less reliable at larger redshifts. We do not apply a minimum redshift cut; 99\% of galaxies in the sample have $z>0.024$.

The hCOSMOS sample is color-selected to target the quiescent galaxy population efficiently with a color range broad enough to achieve high completeness for the quiescent population down to the limiting magnitude \cite[Damjanov et al., In Prep]{Damjanov2015b}. 

The observed-frame colors of galaxies vary with redshift due to redshifting of the SED. At larger redshifts, progressively bluer galaxies satisfy the color selection criteria. Figure \ref{fig:color}A and \ref{fig:color}B show the $g-r$ and $r-i$ rest-frame colors as a function of redshift for the hCOSMOS sample. The red data points in the figure are galaxies with $D_{n}4000 > 1.5$. Based on the $D_{n}4000$ distribution, we derive empirical color-cuts which we apply to the hCOSMOS and SDSS samples. The quiescent galaxy sample has $M_g - M_r > 0.7$, $M_r - M_i > 0.3$ and $D_{n}4000 > 1.5$. We apply these criteria to both samples to select quiescent galaxies consistently. %Our selection is designed to predominately select galaxies where the stellar kinematics are dominated by random stellar motions.

\begin{figure}
\begin{center}
\includegraphics[width=\columnwidth]{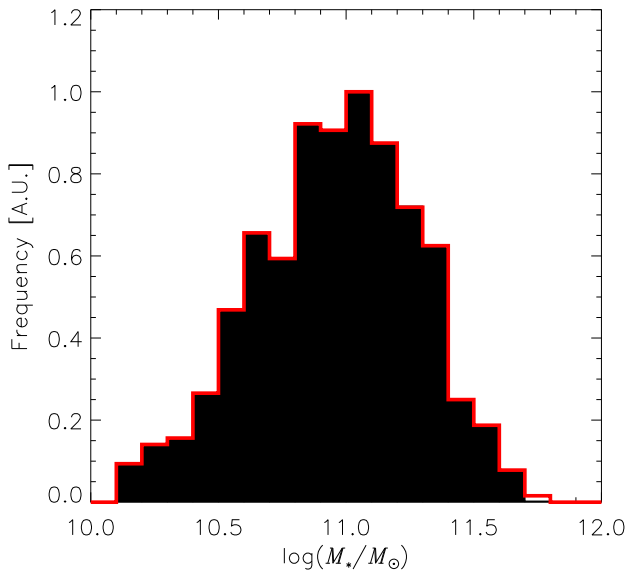}
\end{center}
\caption{Stellar mass distribution of the selected hCOSMOS (red histogram) and SDSS (black filled histogram) samples.}
\label{fig:mass}
\end{figure}

Not all hCOSMOS spectra are of sufficient quality to yield reliable measurements of velocity dispersion. We select galaxies with velocity dispersions measured at a signal-to-noise (S/N) $>3$ and $\chi_{reduced}^2 < 2$. The SDSS spectra are typically higher S/N than the hCOSMOS sample and thus velocity dispersions for the SDSS data are measured down to significantly lower stellar masses. 

To consistently compare the two samples, we randomly select a subset of the SDSS data to match the stellar mass distribution of the hCOSMOS sample. The data are matched in 0.05 dex bins of stellar mass. Figure \ref{fig:mass} shows a histogram of the stellar mass distribution for the two samples. Because we have randomly selected a subset of galaxies as function of stellar mass, the distribution of the other properties of the sample (e.g., velocity dispersion and size) are subject to shot noise. Throughout the remainder of this work, unless otherwise stated, we take an \emph{average} SDSS distribution by combining 1000 randomly selected samples. Any statistical errors quoted account for this oversampling and reflect the statistical errors of a single, randomly selected distribution.

\begin{figure*}
\begin{center}
\includegraphics[width=2\columnwidth]{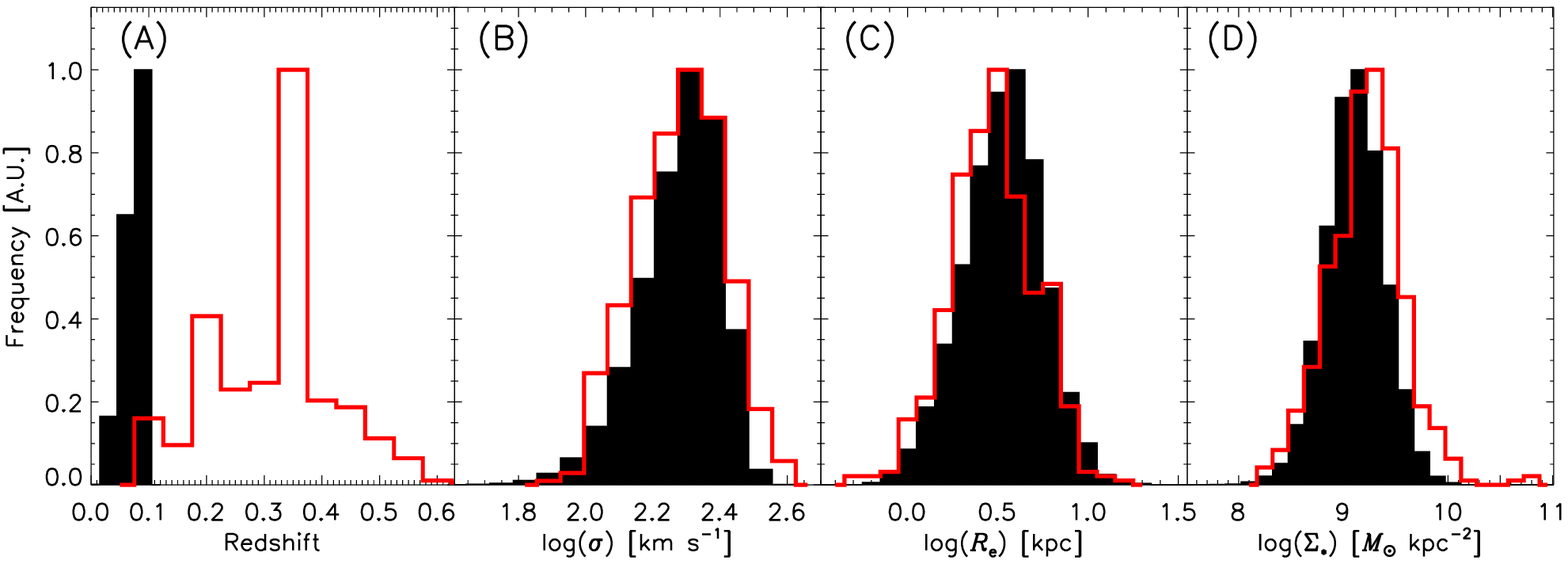}
\end{center}
\caption{Distribution of (A) redshift, (B) velocity dispersion, (C) effective radii and (D) stellar mass surface density. The red and filled black histograms are for the hCOSMOS and SDSS selected samples, respectively.}
\label{fig:hist}
\end{figure*}

Figure \ref{fig:hist} shows a histogram of the physical properties the two samples. The hCOSMOS and SDSS samples are comprised of 509\footnote{Our hCOSMOS survey bright magnitude limit is the magnitude limit of SDSS. We include 45 galaxies in the hCOSMOS sample that are part of the SDSS survey.} and 4970 galaxies, respectively. The positions, redshifts along with the FP properties of galaxies in the hCOSMOS sample are given in Table \ref{tab:prop}.

\subsection{Systematics}

Figure \ref{fig:hist} shows that the velocity dispersion and effective radius distributions of the SDSS and hCOSMOS samples differ. We note that the samples are selected to have the same stellar mass distribution. Thus, differences in the stellar mass density distribution seen in Figure \ref{fig:hist}D reflect only the differences in the effective radius distribution. We show that the differences in Figure \ref{fig:hist}B and \ref{fig:hist}C are not as significant as a direct comparison suggests. Rather, differences in the error distributions and systematic effects contribute significantly to the observed differences in Figure \ref{fig:hist}.
 
Figure \ref{fig:hist}B shows that the velocity dispersion distribution of the SDSS sample is narrower and has a tail at small velocity dispersions. The median velocity dispersion error of the SDSS and hCOSMOS sample is 6 and 30 km s$^{-1}$, respectively. Larger errors in the hCOSMOS velocity dispersions artificially broadens the velocity dispersion distribution. To assess the impact of the differences in the error distribution, we add random artificial errors to the SDSS velocity dispersions so that the simulated SDSS error distribution reproduces the hCOSMOS error distribution. For each object in the SDSS sample we calculate a new error, $\delta_{new, i}$, by randomly drawing an error, $\delta_{Hecto,i}$, from the hCOSMOS distribution in bins of velocity dispersion. Binning the errors is required because the velocity dispersion errors are correlated with the measurement. The subscript $i$ denotes that this procedure is done for each galaxy in the SDSS sample independently. We assume $\delta_{new, i}$ is independent of the true SDSS error, $\delta_{SDSS, i}$. To reproduce the hCOSMOS error distribution, we add this new error in quadrature and require that it satisfy the relation 
 \begin{equation}
 \delta_{new, i}^2 = \delta_{Hecto, i}^2 - \delta_{SDSS, i}^2.
 \end{equation}
In short, $\delta_{new,i}$ is the randomly generated error which, when added in quadrature to the true SDSS velocity dispersion error, yields a new distribution of SDSS errors that replicates the hCOSMOS velocity dispersion error distribution. We emphasize that the match in the velocity dispersion distributions of the two samples is solely a consequence of matching the \emph{error} distributions of the velocity dispersion measurements.

\begin{figure*}
\begin{center}
\includegraphics[width=1.5\columnwidth]{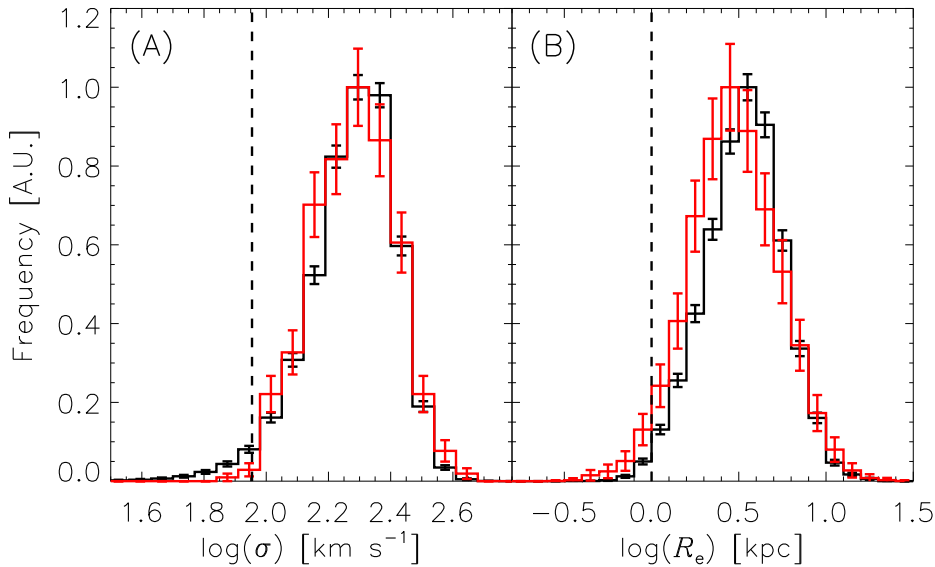}
\end{center}
\caption{The (A) velocity dispersion and (B) size distribution of the SDSS (black) and hCOSMOS (red) samples. Artificial errors have been added to (A) the velocity dispersions of SDSS galaxies and (B) the sizes of the hCOSMOS sample as described in the text. The error bars represent Poisson uncertainties. The dashed lines show (A) the Hectospec resolution limit and (B) the minimum size that can be reliably measured for a galaxy at the median redshift of the SDSS sample ($z \sim 0.08$) assuming the typical $1''\!\!.2$ seeing.}
\label{fig:new_hist}
\end{figure*}

We add the artificial error, $\delta_{new,i}$ to each SDSS velocity dispersion measurement by randomly drawing from a Gaussian with standard deviation, $\delta_{new,i}$. Figure \ref{fig:new_hist}A shows that after the errors are added to the velocity dispersion distribution of the SDSS sample, the two distributions are much more consistent over most of the range in velocity dispersion. Clear differences are present in the two distributions at small velocity dispersion. This systematic effect results from the higher resolution of the SDSS spectrograph. Figure \ref{fig:new_hist}A shows the minimum velocity dispersion we can reliably measure for hCOSMOS, set by the Hectospec resolution. We can not account for the limiting resolution of the two spectrographs by adding random errors to the SDSS velocity dispersions. 

In this analysis we have assumed that the errors are normally distributed and that velocity dispersions are independent of other physical properties. These assumptions are probably not valid in detail and likely contribute to the lingering differences in the two velocity dispersion distributions.

We perform a similar analysis to assess differences in the effective radius distributions. In this case however, the errors are significantly larger for the SDSS; we add artificial errors to the hCOSMOS size distribution. To derive the artificial errors, we assume that the differences in the effective radii measurements seen in the Figure \ref{fig:size}A are due solely to errors in the SDSS measurements. The \emph{HST} measurements are of significantly higher quality and thus observational uncertainties in these measurements are significantly smaller than the SDSS measurements. 

We randomly sample the distribution of the difference in the two measurements and add this difference to the \emph{HST} measurement. Figure \ref{fig:new_hist}B shows that the two effective size distributions are more consistent after we add artificial errors to the \emph{HST} effective radii. At intermediate sizes the hCOSMOS galaxies are systematically smaller by $\sim0.05$ dex. We attribute this difference to real size evolution in the population. However, we emphasize that this conclusion remains tentative; potentially unaccounted for systematic effects related to the seeing and spectroscopic completeness may also produce similar trends \citep{Taylor2010, Carollo2013}. Figure \ref{fig:new_hist}B shows the minimum size that can be reliably measured with SDSS photometry at the median redshift of the SDSS data. A tail in the \emph{HST} size distribution is still present at small radii (8\% and 3\% of galaxies have radii smaller than 1 kpc for the hCOSMOS and SDSS samples, respectively). The tail likely results from the limiting resolution of the SDSS photometry, a systematic effect which is not accounted for by the procedure of simply adding artificial error. 

Our error analysis suggests that galaxies in the SDSS and hCOSMOS samples are in fact drawn from the same velocity dispersion distribution. We emphasize that this agreement would be obscured without the detailed analysis of the error distribution and sample selection. Intermediate size SDSS galaxies are $\sim0.05$ dex larger than hCOSMOS galaxies, an effect we tentatively attribute to size evolution. For the following analysis, errors are added to the velocity dispersions of SDSS galaxies and sizes of hCOSMOS galaxies as described above and objects with log$(\sigma)  < 1.9$ and log$(R_e)  <  0$ are excluded. However, in Figures \ref{fig:mfp} and \ref{fig:mfp_fp} we display the original data.

\section{The Mass Fundamental Plane}

\begin{figure*}
\begin{center}
\includegraphics[width=1.5\columnwidth]{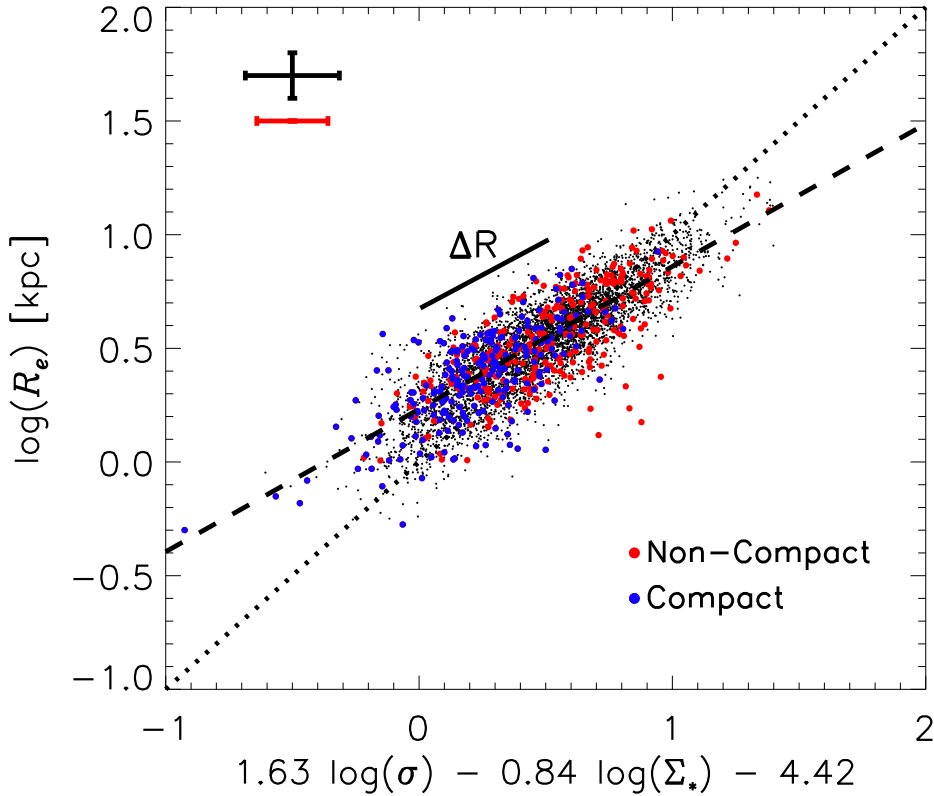}
\end{center}
\caption{Stellar mass fundamental plane for galaxies at $z<0.6$. The black dots are SDSS galaxies and the blue and red points are galaxies in the hCOSMOS sample. The black and red error bars are the median errors for the SDSS and hCOSMOS samples, respectively. For display, we have not added artificial errors to the data and have not applied minimum velocity dispersion and effective radii cuts as described in the text. The dotted line is the \citet{Hyde2009b} relation and the dashed line is the projection of the best-fit plane derived from the SDSS data.}
\label{fig:mfp}
\end{figure*}
 
We show the MFP for the two samples in Figure \ref{fig:mfp}. The data are plotted as a function of the orthogonal $r-$band MFP fit parameters given in \citet{Hyde2009b}. The dashed line is the \citet{Hyde2009b} relation. The data scatter off this relation. The solid line denoted by $\Delta R$ is a vector indicating how galaxies move on this plot if the radius changes while all other physical properties remain fixed. The size measurements used by \citet{Hyde2009b} to derive the FP and MFP are from the SDSS photometric pipeline and are  inconsistent with the \emph{HST} size measurements of S07 (see Figure \ref{fig:size}B). The data points scatter off the \citet{Hyde2009b} relation parallel to the $\Delta R$ vector suggesting that systematic differences in the size measurements contribute to the difference in the two relations. Additionally, selection effects likely contribute to the differences in the two relations.

Compact galaxies are shown by the blue dots in Figure \ref{fig:mfp}. These galaxies have smaller sizes and larger stellar mass densities. Thus, by definition they populate the lower left region of Figure \ref{fig:mfp}. 
 
%\begin{deluxetable*}{ccccc}
%\tablewidth{350pt}
%\tablecaption{Principal Components}
%\tablehead{\colhead{Component} & \colhead{log$(\Sigma_\ast)$} &\colhead{log$(\sigma)$} & \colhead{log$(R_e)$} & \colhead{Variance [\%]} }
%\startdata
%\noalign{\smallskip}
%\multicolumn{5}{c}{SDSS}\\
%\hline
%\noalign{\smallskip}
%PCA 1 &-0.907$\pm$0.004 & 0.351$\pm$0.023 & 0.970$\pm$0.001 & 62.9\%  \\
%PCA 2 &0.3550$\pm$0.012 & 0.932$\pm$0.009 & -0.00$\pm$0.011 & 33.2\% \\
%PCA 3 &0.2270$\pm$0.003 & -0.08$\pm$0.003 & 0.243$\pm$0.003 & 3.9\% \\
%\hline
%\noalign{\smallskip}
%\multicolumn{5}{c}{hCOSMOS}\\
%\noalign{\smallskip}
%\hline
%\noalign{\smallskip}
%PCA 1 &-0.961$\pm$0.004 & -0.21$\pm$0.096 & 0.919$\pm$0.016 & 60.4\% \\
%PCA 2 &-0.073$\pm$0.053 & -0.97$\pm$0.021 & -0.30$\pm$0.052 & 34.7\% \\
%PCA 3 &-0.266$\pm$0.012 & 0.099$\pm$0.009 & -0.25$\pm$0.011 & 4.9\% \\
%\enddata
%\label{tab:fit}
%\tablecomments{}
%\end{deluxetable*}

We perform a principal component analysis (PCA) on the MFP data, i.e. $\sigma$, $R_e$ and $\Sigma_\ast$, to find the best-fit plane determined from minimizing orthogonal residuals. PCA is a non-parametric method which finds the set of orthogonal vectors which best account for variance in multivariate data \citep{Shlens2014}. The first component accounts for the maximum amount of variance in the data and each successive component accounts for the maximum amount of remaining variance under the condition that all components are orthogonal. The first two principal components form a plane that accounts for 94\% of the variance in the SDSS and hCOSMOS samples; as expected for three dimensional data distributed in plane. By definition, the third principal component is the vector normal to the plane containing the first two components; a plane can be defined by its normal vector. Thus, the third principal component defines the best-fit MFP determined from minimization of \emph{orthogonal} residuals. We consider minimization of orthogonal residuals more robust than a parametric model fit because none of the variables can be considered independent and all the data used in defining the MFP carry observational uncertainties.

We normalize the third principal component vector to the effective radius coefficient. The normal vectors defining the MFP for the SDSS and hCOSMOS samples are 
\begin{equation*}
(-0.955 \pm 0.017, 0.486 \pm 0.015, -1.000 \pm 0.018) 
\nonumber
\end{equation*}
and 
\begin{equation*}
(-1.096 \pm 0.069, 0.519 \pm 0.051, -1.000 \pm 0.063),
\end{equation*}
respectively. A projection of the best-fit plane derived from the SDSS data is shown in Figure \ref{fig:mfp}. The differences in the parameters are not statistically significant ($<2\sigma$). The errors on the vector components are bootstrapped. These vectors correspond to the equation
\begin{equation}
a\mathrm{log}(\Sigma_\ast) + b\mathrm{log}(\sigma) + c\mathrm{log}(R_e) + d = 0, 
\label{eq:mfp_fit}
\end{equation}
where $(a,b,c)$ are the normal vectors given above. The orientation of the best-fit MFP is consistent for the two samples. We calculate the zero-point of the MFP which we define as 
\begin{equation}
d = -a\mean{ \mathrm{log}(\Sigma_\ast) } - b\mean{ \mathrm{log}(\sigma)} - c \mean{ \mathrm{log}(R_e) }. 
\end{equation}
Here, $\mean{}$ denotes the mean of the distribution. The zero-point is covariant with parameters $a$, $b$ and $c$. We calculate the zero-point for both samples using the SDSS parameters. The zero-point is $8.118 \pm0.003$ and $8.160 \pm 0.011$ for SDSS and hCOSMOS samples, respectively\footnote{If we adopt the best-fit parameters for each sample and account for the covariance between the slope parameters and the zero-point, we measure a zero-point of $8.12\pm0.16$ and $9.39\pm0.66$ for the SDSS and hCOSMOS sample, respectively.} Thus, the orientation of the best-fit MFP determined from minimizing orthogonal residuals is consistent for the hCOSMOS and SDSS samples and the zero-point significantly ($3.5\sigma$) evolves by $0.042\pm0.011$ dex.

\begin{figure*}
\begin{center}
\includegraphics[width=2\columnwidth]{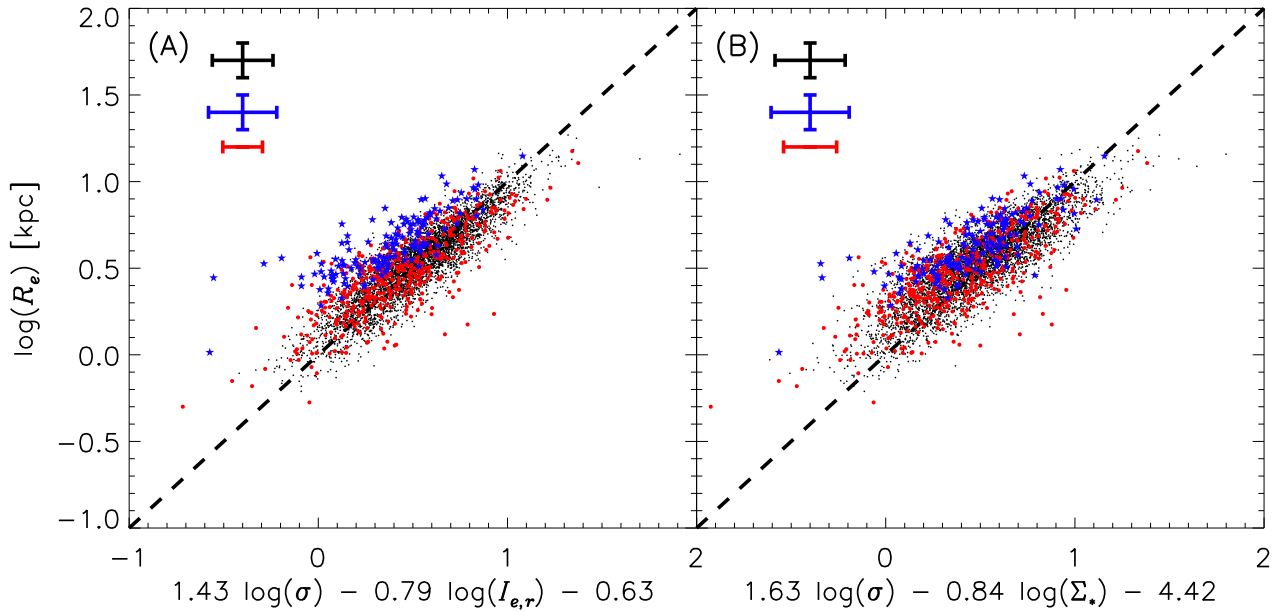}
\end{center}
\caption{The (A) FP and (B) MFP for the SDSS (black points), hCOSMOS (red dots) and Z15 (blue stars) samples. The black, red and blue error bars are the median errors for the SDSS, hCOSMOS and Z15 samples, respectively. The dashed line is the \citet{Hyde2009b} relation.}
\label{fig:mfp_fp}
\end{figure*}

\section{Discussion}

The FP and MFP have a known dependence on selection effects \citep{Hyde2009b}. Attempts to measure evolution in the structural and dynamical properties of galaxies thus require careful control of selection bias. A comparison of Figure \ref{fig:hist} and \ref{fig:new_hist} demonstrates that systematic effects must also be considered when examining the distribution of galaxy properties. These issues are especially serious when different facilities and methods are used to study galaxies.

We examine the evolution of the MFP by comparing our sample at $z<0.6$ to a local ``control'' sample. We apply the same quiescent galaxy selection criteria to both samples. We match the stellar mass distributions of the two samples by randomly selecting a subset of galaxies from the local parent sample. We account for systematic effects related to the quality of the measurements by adding simulated errors to the data requiring that the subsequent size and velocity dispersion error distributions of the two samples match. We account for major systematic effects by matching the effects across the two samples. After matching the systematic effects across the two samples, the primary difference between the two samples is the redshift distribution of galaxies. Thus, a comparison of the MFP for the two samples, which we derive using a non-parametric principle component analysis, provides a robust measure of the redshift evolution of the MFP. We emphasize that we derive a \emph{relative}$-$not an absolute$-$measurement of the parameters of the MFP and their evolution. In contrast, other works apply analytic corrections to derive an estimate of the absolute parameters of the MFP \citep[e.g.,][]{Hyde2009b, Saulder2013}; we only measure whether there is relative evolution in these parameters. 

We show that the structural and dynamical properties of massive quiescent galaxies at $z<0.6$ have no strong dependence on redshift. \citet{Bezanson2013} also find a very mild redshift evolution of the zero-point of the MFP ($\sim0.02 \pm 0.01$ dex lower at $z\sim0.6$). However, because of their small sample size, \citet{Bezanson2013} have to assume that the orientation of the MFP does not evolve with redshift. Our larger sample size allows measurement of both the orientation and zero-point of the MFP. The orientation of the MFP for massive quiescent galaxies does not evolve significantly with redshift for $z<0.6$ \citep[also see][]{Holden2010}. We measure a small, but significant evolution in the zero-point ($0.04$ dex) which may be due to size evolution in the galaxy population though this conclusion remains tentative because of potential systematic effects not accounted for in this study.

A number of studies indicate that the average size of the quiescent galaxy population increases with time \citep[e.g.,][]{Daddi2005}. Various scenarios have been proposed to account for this size increase. Either individual quiescent galaxies grow after they are quenched \citep[e.g.,][]{Newman2012}, or galaxies that become quiescent at later times are larger(i.e. progenitor bias), thus increasing the average size of the population as a function of time \citep[e.g.,][]{Carollo2013}; both processes may be relevant. In the case of individual quiescent galaxy growth, potential mechanisms include major mergers, minor mergers and accretion \citep[e.g.,][]{Naab2009} or adiabatic expansion driven by feedback \citep{Fan2008}. Simulations suggest that minor mergers do not significantly change the stellar velocity dispersion, whereas major mergers and adiabatic driven expansion substantially increase and decrease the stellar velocity dispersion, respectively \citep{Hopkins2010}. Our results can not discriminate between growth of individual quiescent galaxies or progenitor bias as the basis for the increase in the average size of the quiescent galaxy population with time. However, our results do indicate a small amount of size growth with no change in the velocity dispersion distribution at $z<0.6$. Thus, \emph{if} individual galaxies do grow with time, the galaxy size evolution we measure is most consistent with a minor merger driven growth scenario.

Our data indicate that the average growth in the quiescent galaxy population is $\lesssim 0.1$ dex. Our estimates serve as an upper limit owing to potentially unaccounted for systematic uncertainties. \cite{vanderWel2014} measure the growth of quiescent galaxies at $z<3$ using deep HST imaging. Based on their preferred model fit to the data, we calculate an average growth of 0.08 dex for quiescent galaxies between the $z\sim 0.35$ and $z\sim0.07$ (the median redshifts of the hCOSMSO and SDSS samples, respectively). Thus, our conclusions regarding the growth of quiescent galaxies are consistent with the size evolution reported by \citet{vanderWel2014}.

Offsets from the FP correlate with galaxy properties and redshift \citep[e.g.,][Z15]{Treu2005, vanDokkum2007, Holden2010, Bezanson2013, vandeSande2014}. These offsets are typically attributed to variations in the M/L ratios of galaxies. To explore this issue, we compare the Z15 sample with the samples used in this study. Z15 examine the FP of intermediate redshift quiescent galaxies. Figure \ref{fig:mfp_fp} shows that the Z15 data are offset from the data examined in this study. The Z15 sample are $z>0.2$ galaxies observed as part of the SDSS. Because of selection effects present in a magnitude limited survey like the SDSS, the Z15 sample is strongly biased towards bright, high surface brightness galaxies; the Z15 sample represent the tail of the luminosity and surface brightness distribution of galaxies at the redshifts probed by the sample. Because of selection effects, we expect galaxies in the Z15 sample to have lower M/L ratios as compared to similar galaxies in the hCOSMOS and SDSS samples we examine in this study. Thus, we expect that the FP offsets should be significantly larger than offsets from the MFP. The smaller MFP offsets are consistent with the fact that stellar mass estimates, in principle, account for variations in the M/L ratio. 

Despite accounting for the M/L ratio, the Z15 data are offset from the MFP. The zero-point of the MFP ($d$ in Equation \ref{eq:mfp_fit}) measured from the Z15 data is $8.36\pm0.02$. This is significantly different from the zero-point determined from the data in this study. Several systematic effects could contribute to the offset. The stellar masses and velocity dispersions of the Z15 sample are calculated using different techniques and it is beyond the scope of this paper to perform a detailed comparison. Furthermore, objects in the Z15 sample are extreme because selection is strongly biased towards the highest surface brightness objects; the hCOSMOS and SDSS samples in this study are not subject to such a strong selection bias.

%Objects in the Z15 sample are extreme in several physical properties. The Z15 sample is significantly bluer than the data examined in this study. and therefore M/L ratio variations are significantly larger \citep[e.g.][]{Geller2014}. A systematic overestimate in the M/L ratio would move galaxies to the left in Figure \ref{fig:mfp_fp}B and could account for some of the offset. Systematics differences in estimates of other parameters may also contribute. 

The comparison of the Z15 sample with the hCOSMOS and SDSS samples highlights the impact of selection and systematic effects. Based on Figure \ref{fig:mfp_fp}, we conclude that M/L ratio variations account for much of the offset between the Z15 FP and the SDSS and hCOSMOS sample. However, the residual offsets observed in Figure \ref{fig:mfp_fp}B highlight the necessity for strict control of selection effects and systematics related to measurement procedures. Figure \ref{fig:mfp_fp} indeed shows how a lack of control of these effects could lead to spurious trends with redshift.

Real variations in the M/L ratio contribute to the scatter in the FP (e.g., Z15). In this context, the smaller scatter in the FP as compared to the MFP in Figure \ref{fig:mfp_fp} may be unexpected. The RMS orthogonal scatter in the FP and MFP for the SDSS sample is 0.14 and 0.17 dex, respectively; for the hCOSMOS sample it is 0.21 and 0.24 dex, respectively. By comparing stellar masses with absolute magnitudes, we empirically determine that the scatter in the $r$-band M/L ratio is 0.10 dex for both samples used in this study. This additional scatter of 0.10 dex added in quadrature to the scatter in the FP accounts for the 0.03 dex greater scatter in the MFP. The SDSS and hCOSMOS sample of galaxies are red-selected with $D_n4000 > 1.5$. The M/L ratio is nearly constant for these objects \citep[e.g.,][]{Geller2014}. Thus, the larger scatter in the MFP as compared to the FP suggests that measurement uncertainties associated with the stellar mass determination are greater than real variations in the M/L ratio and uncertainties in the $k-$correction for galaxies in the two samples. Because of the very red selection, no redshift evolution correction to the luminosity is necessary for galaxies in the redshift range probed in this study. Thus, samples selected according to the approach outlined here may prove useful for investigating evolution of the stellar mass in relation to the dynamical mass since uncertainties related to M/L ratio are small and luminosities can be taken as direct proxies for stellar mass.

It appears that compact galaxies are not a special class of objects. Z15 show that passive evolution alone brings the massive compact quiescent galaxies observed in SDSS at intermediate redshifts onto the local FP relation. Thus, Z15 conclude that massive compact galaxies are the high mass, high surface brightness tail of the normal galaxy distribution. Figures \ref{fig:mfp} and \ref{fig:mfp_fp} support this conclusion. Both the MFP and FP of compact quiescent galaxies out to $z\sim0.6$ are consistent with the local relation. This conclusion regarding the nature of compact galaxies is also supported by recent analysis of the environments of compact galaxies \citep{Damjanov2015b}. Compact quiescent galaxies inhabit the same environments as similarly massive quiescent galaxies. Recent cosmological hydrodynamical simulations \citep{Wellons2015a} and dynamical studies of intermediate redshift compact galaxies \citep{Saulder2015} also suggest that compact galaxies are not a special class of objects.

Z15 show that any size growth in the compact quiescent galaxy population requires that they remain on the FP as they evolve, i.e. they do not deviate strongly from virial equilibrium. We confirm this constraint with our larger sample of galaxies at $z<0.6$. After controlling for selection and systematic effects, we find that massive quiescent galaxies in the hCOSMOS sample are slightly smaller compared with similar galaxies in the local universe (Figure \ref{fig:new_hist}). Moreover, the fraction of very small galaxies in the hCOSMOS sample is larger than the SDSS sample. The observed difference serves as an upper limit for the amount of size growth in the compact galaxy population. For the redshifts probed by this study we estimate that compact quiescent galaxies may grow by $\lesssim0.1$ dex. However, the number density of compact quiescent galaxies remains nearly constant at $z<1$ \citep{Carollo2013, Damjanov2015a}; thus any significant growth in the quiescent compact galaxy population must be balanced by the production of new compact galaxies at $z<0.6$.

\section{Summary and Conclusions}

We examine the stellar mass fundamental plane for massive quiescent galaxies at $z<0.6$ and compare the result to the local stellar mass fundamental plane from SDSS. We examine the relation between stellar mass density, velocity dispersion and size using a principle component analysis. The orientation of the stellar mass fundamental plane is independent of redshift and the zero-point evolves by $-0.042 \pm 0.011$ dex. We tentatively attribute the zero-point evolution to size evolution of the population. However, this small evolution may be due to unaccounted for systematic effects. If the observed average size growth is due to growth of individual quiescent galaxies, our data is most consistent with minor merger driven growth scenario. Our analysis demonstrates that both systematic and selection effects must be accounted for when examining the structural and dynamical properties of galaxies and the evolution of these properties with redshift. 

The mass-to-light ratios of quiescent galaxies are nearly constant. Thus luminosities of quiescent galaxies selected on the basis of their photometric and spectroscopic properties can be taken as a proxy for stellar mass without invoking uncertainties associated with stellar mass determinations. Thus, the luminosities of the quiescent population of galaxies (if carefully selected) provide additional constraints for examining the redshift evolution of baryonic versus dynamical mass within galaxies. 

We identify compact galaxies in our sample and show that their dynamical and structural properties are consistent with being drawn from the same distribution as the non-compact quiescent galaxy population. Thus, either compact galaxies do not undergo significant growth in size or if they do, they are replaced by newly formed compact quiescent galaxies. We confirm previous results suggesting that compact galaxies are not a special class of objects but the tail of the size and mass distribution of the normal quiescent galaxy population.

\acknowledgements

HJZ gratefully acknowledges the generous support of the Clay Postdoctoral Fellowship. ID is supported by the Harvard College Observatory Menzel Fellowship and the Natural Sciences and Engineering Research Council of Canada Postdoctoral Fellowship (NSERC PDF-421224-2012). MJG is supported by the Smithsonian Institution. This research has made use of NASA's Astrophysics Data System Bibliographic Services.

Funding for SDSS-III has been provided by the Alfred P. Sloan Foundation, the Participating Institutions, the National Science Foundation, and the U.S. Department of Energy Office of Science. The SDSS-III web site is http://www.sdss3.org/.

SDSS-III is managed by the Astrophysical Research Consortium for the Participating Institutions of the SDSS-III Collaboration including the University of Arizona, the Brazilian Participation Group, Brookhaven National Laboratory, University of Cambridge, Carnegie Mellon University, University of Florida, the French Participation Group, the German Participation Group, Harvard University, the Instituto de Astrofisica de Canarias, the Michigan State/Notre Dame/JINA Participation Group, Johns Hopkins University, Lawrence Berkeley National Laboratory, Max Planck Institute for Astrophysics, Max Planck Institute for Extraterrestrial Physics, New Mexico State University, New York University, Ohio State University, Pennsylvania State University, University of Portsmouth, Princeton University, the Spanish Participation Group, University of Tokyo, University of Utah, Vanderbilt University, University of Virginia, University of Washington, and Yale University.

\bibliographystyle{apj}
\bibliography{/Users/jabran/Documents/latex/metallicity}

\begin{thebibliography}{76}
\expandafter\ifx\csname natexlab\endcsname\relax\def\natexlab#1{#1}\fi

\bibitem[{{Alam} {et~al.}(2015){Alam}, {Albareti}, {Allende Prieto}, {Anders},
  {Anderson}, {Anderton}, {Andrews}, {Armengaud}, {Aubourg}, {Bailey}, \&
  et~al.}]{Alam2015}
{Alam}, S., {et~al.} 2015, \apjs, 219, 12

\bibitem[{{Arnouts} {et~al.}(1999){Arnouts}, {Cristiani}, {Moscardini},
  {Matarrese}, {Lucchin}, {Fontana}, \& {Giallongo}}]{Arnouts1999}
{Arnouts}, S., {Cristiani}, S., {Moscardini}, L., {Matarrese}, S., {Lucchin},
  F., {Fontana}, A., \& {Giallongo}, E. 1999, \mnras, 310, 540

\bibitem[{{Balogh} {et~al.}(1999){Balogh}, {Morris}, {Yee}, {Carlberg}, \&
  {Ellingson}}]{Balogh1999}
{Balogh}, M.~L., {Morris}, S.~L., {Yee}, H.~K.~C., {Carlberg}, R.~G., \&
  {Ellingson}, E. 1999, \apj, 527, 54

\bibitem[{{Barro} {et~al.}(2013){Barro}, {Faber}, {P{\'e}rez-Gonz{\'a}lez},
  {Koo}, {Williams}, {Kocevski}, {Trump}, {Mozena}, {McGrath}, {van der Wel},
  {Wuyts}, {Bell}, {Croton}, {Ceverino}, {Dekel}, {Ashby}, {Cheung},
  {Ferguson}, {Fontana}, {Fang}, {Giavalisco}, {Grogin}, {Guo}, {Hathi},
  {Hopkins}, {Huang}, {Koekemoer}, {Kartaltepe}, {Lee}, {Newman}, {Porter},
  {Primack}, {Ryan}, {Rosario}, {Somerville}, {Salvato}, \& {Hsu}}]{Barro2013}
{Barro}, G., {et~al.} 2013, \apj, 765, 104

\bibitem[{{Bernardi} {et~al.}(2003){Bernardi}, {Sheth}, {Annis}, {Burles},
  {Eisenstein}, {Finkbeiner}, {Hogg}, {Lupton}, {Schlegel}, {SubbaRao},
  {Bahcall}, {Blakeslee}, {Brinkmann}, {Castander}, {Connolly}, {Csabai},
  {Doi}, {Fukugita}, {Frieman}, {Heckman}, {Hennessy}, {Ivezi{\'c}}, {Knapp},
  {Lamb}, {McKay}, {Munn}, {Nichol}, {Okamura}, {Schneider}, {Thakar}, \&
  {York}}]{Bernardi2003}
{Bernardi}, M., {et~al.} 2003, \aj, 125, 1866

\bibitem[{{Bezanson} {et~al.}(2013){Bezanson}, {van Dokkum}, {van de Sande},
  {Franx}, {Leja}, \& {Kriek}}]{Bezanson2013}
{Bezanson}, R., {van Dokkum}, P.~G., {van de Sande}, J., {Franx}, M., {Leja},
  J., \& {Kriek}, M. 2013, \apjl, 779, L21

\bibitem[{{Blanton} {et~al.}(2005{\natexlab{a}}){Blanton}, {Eisenstein},
  {Hogg}, {Schlegel}, \& {Brinkmann}}]{Blanton2005b}
{Blanton}, M.~R., {Eisenstein}, D., {Hogg}, D.~W., {Schlegel}, D.~J., \&
  {Brinkmann}, J. 2005{\natexlab{a}}, \apj, 629, 143

\bibitem[{{Blanton} {et~al.}(2005{\natexlab{b}}){Blanton}, {Schlegel},
  {Strauss}, {Brinkmann}, {Finkbeiner}, {Fukugita}, {Gunn}, {Hogg},
  {Ivezi{\'c}}, {Knapp}, {Lupton}, {Munn}, {Schneider}, {Tegmark}, \&
  {Zehavi}}]{Blanton2005a}
{Blanton}, M.~R., {et~al.} 2005{\natexlab{b}}, \aj, 129, 2562

\bibitem[{{Bolton} {et~al.}(2008){Bolton}, {Treu}, {Koopmans}, {Gavazzi},
  {Moustakas}, {Burles}, {Schlegel}, \& {Wayth}}]{Bolton2008}
{Bolton}, A.~S., {Treu}, T., {Koopmans}, L.~V.~E., {Gavazzi}, R., {Moustakas},
  L.~A., {Burles}, S., {Schlegel}, D.~J., \& {Wayth}, R. 2008, \apj, 684, 248

\bibitem[{{Bruzual} \& {Charlot}(2003)}]{Bruzual2003}
{Bruzual}, G., \& {Charlot}, S. 2003, \mnras, 344, 1000

\bibitem[{{Buitrago} {et~al.}(2008){Buitrago}, {Trujillo}, {Conselice},
  {Bouwens}, {Dickinson}, \& {Yan}}]{Buitrago2008}
{Buitrago}, F., {Trujillo}, I., {Conselice}, C.~J., {Bouwens}, R.~J.,
  {Dickinson}, M., \& {Yan}, H. 2008, \apjl, 687, L61

\bibitem[{{Calzetti} {et~al.}(2000){Calzetti}, {Armus}, {Bohlin}, {Kinney},
  {Koornneef}, \& {Storchi-Bergmann}}]{Calzetti2000}
{Calzetti}, D., {Armus}, L., {Bohlin}, R.~C., {Kinney}, A.~L., {Koornneef}, J.,
  \& {Storchi-Bergmann}, T. 2000, \apj, 533, 682

\bibitem[{{Cappellari} {et~al.}(2006){Cappellari}, {Bacon}, {Bureau}, {Damen},
  {Davies}, {de Zeeuw}, {Emsellem}, {Falc{\'o}n-Barroso}, {Krajnovi{\'c}},
  {Kuntschner}, {McDermid}, {Peletier}, {Sarzi}, {van den Bosch}, \& {van de
  Ven}}]{Cappellari2006}
{Cappellari}, M., {et~al.} 2006, \mnras, 366, 1126

\bibitem[{{Cappellari} \& {Emsellem}(2004)}]{Cappellari2004}
{Cappellari}, M., \& {Emsellem}, E. 2004, \pasp, 116, 138

\bibitem[{{Carollo} {et~al.}(2013){Carollo}, {Bschorr}, {Renzini}, {Lilly},
  {Capak}, {Cibinel}, {Ilbert}, {Onodera}, {Scoville}, {Cameron}, {Mobasher},
  {Sanders}, \& {Taniguchi}}]{Carollo2013}
{Carollo}, C.~M., {et~al.} 2013, \apj, 773, 112

\bibitem[{{Cassata} {et~al.}(2013){Cassata}, {Giavalisco}, {Williams}, {Guo},
  {Lee}, {Renzini}, {Ferguson}, {Faber}, {Barro}, {McIntosh}, {Lu}, {Bell},
  {Koo}, {Papovich}, {Ryan}, {Conselice}, {Grogin}, {Koekemoer}, \&
  {Hathi}}]{Cassata2013}
{Cassata}, P., {et~al.} 2013, \apj, 775, 106

\bibitem[{{Chabrier}(2003)}]{Chabrier2003}
{Chabrier}, G. 2003, \pasp, 115, 763

\bibitem[{{Daddi} {et~al.}(2005){Daddi}, {Renzini}, {Pirzkal}, {Cimatti},
  {Malhotra}, {Stiavelli}, {Xu}, {Pasquali}, {Rhoads}, {Brusa}, {di Serego
  Alighieri}, {Ferguson}, {Koekemoer}, {Moustakas}, {Panagia}, \&
  {Windhorst}}]{Daddi2005}
{Daddi}, E., {et~al.} 2005, \apj, 626, 680

\bibitem[{{Damjanov} {et~al.}(2011){Damjanov}, {Abraham}, {Glazebrook},
  {McCarthy}, {Caris}, {Carlberg}, {Chen}, {Crampton}, {Green}, {J{\o}rgensen},
  {Juneau}, {Le Borgne}, {Marzke}, {Mentuch}, {Murowinski}, {Roth}, {Savaglio},
  \& {Yan}}]{Damjanov2011}
{Damjanov}, I., {et~al.} 2011, \apjl, 739, L44

\bibitem[{{Damjanov} {et~al.}(2015{\natexlab{a}}){Damjanov}, {Geller}, {Zahid},
  \& {Hwang}}]{Damjanov2015a}
{Damjanov}, I., {Geller}, M.~J., {Zahid}, H.~J., \& {Hwang}, H.~S.
  2015{\natexlab{a}}, \apj, 806, 158

\bibitem[{{Damjanov} {et~al.}(2014){Damjanov}, {Hwang}, {Geller}, \&
  {Chilingarian}}]{Damjanov2014}
{Damjanov}, I., {Hwang}, H.~S., {Geller}, M.~J., \& {Chilingarian}, I. 2014,
  \apj, 793, 39

\bibitem[{{Damjanov} {et~al.}(2015{\natexlab{b}}){Damjanov}, {Zahid}, {Geller},
  \& {Hwang}}]{Damjanov2015b}
{Damjanov}, I., {Zahid}, H.~J., {Geller}, M.~J., \& {Hwang}, H.~S.
  2015{\natexlab{b}}, ArXiv e-prints

\bibitem[{{Djorgovski} \& {Davis}(1987)}]{Djorgovski1987}
{Djorgovski}, S., \& {Davis}, M. 1987, \apj, 313, 59

\bibitem[{{Doi} {et~al.}(2010){Doi}, {Tanaka}, {Fukugita}, {Gunn}, {Yasuda},
  {Ivezi{\'c}}, {Brinkmann}, {de Haars}, {Kleinman}, {Krzesinski}, \& {French
  Leger}}]{Doi2010}
{Doi}, M., {et~al.} 2010, \aj, 139, 1628

\bibitem[{{Dressler} {et~al.}(1987){Dressler}, {Lynden-Bell}, {Burstein},
  {Davies}, {Faber}, {Terlevich}, \& {Wegner}}]{Dressler1987}
{Dressler}, A., {Lynden-Bell}, D., {Burstein}, D., {Davies}, R.~L., {Faber},
  S.~M., {Terlevich}, R., \& {Wegner}, G. 1987, \apj, 313, 42

\bibitem[{{Fabricant} {et~al.}(2013){Fabricant}, {Chilingarian}, {Hwang},
  {Kurtz}, {Geller}, {Del'Antonio}, \& {Rines}}]{Fabricant2013}
{Fabricant}, D., {Chilingarian}, I., {Hwang}, H.~S., {Kurtz}, M.~J., {Geller},
  M.~J., {Del'Antonio}, I.~P., \& {Rines}, K.~J. 2013, \pasp, 125, 1362

\bibitem[{{Fabricant} {et~al.}(2005){Fabricant}, {Fata}, {Roll}, {Hertz},
  {Caldwell}, {Gauron}, {Geary}, {McLeod}, {Szentgyorgyi}, {Zajac}, {Kurtz},
  {Barberis}, {Bergner}, {Brown}, {Conroy}, {Eng}, {Geller}, {Goddard},
  {Honsa}, {Mueller}, {Mink}, {Ordway}, {Tokarz}, {Woods}, {Wyatt}, {Epps}, \&
  {Dell'Antonio}}]{Fabricant2005}
{Fabricant}, D., {et~al.} 2005, \pasp, 117, 1411

\bibitem[{{Fan} {et~al.}(2008){Fan}, {Lapi}, {De Zotti}, \& {Danese}}]{Fan2008}
{Fan}, L., {Lapi}, A., {De Zotti}, G., \& {Danese}, L. 2008, \apjl, 689, L101

\bibitem[{{Geller} {et~al.}(2005){Geller}, {Dell'Antonio}, {Kurtz}, {Ramella},
  {Fabricant}, {Caldwell}, {Tyson}, \& {Wittman}}]{Geller2005}
{Geller}, M.~J., {Dell'Antonio}, I.~P., {Kurtz}, M.~J., {Ramella}, M.,
  {Fabricant}, D.~G., {Caldwell}, N., {Tyson}, J.~A., \& {Wittman}, D. 2005,
  \apjl, 635, L125

\bibitem[{{Geller} {et~al.}(2014){Geller}, {Hwang}, {Fabricant}, {Kurtz},
  {Dell'Antonio}, \& {Zahid}}]{Geller2014}
{Geller}, M.~J., {Hwang}, H.~S., {Fabricant}, D.~G., {Kurtz}, M.~J.,
  {Dell'Antonio}, I.~P., \& {Zahid}, H.~J. 2014, \apjs, 213, 35

\bibitem[{{Holden} {et~al.}(2010){Holden}, {van der Wel}, {Kelson}, {Franx}, \&
  {Illingworth}}]{Holden2010}
{Holden}, B.~P., {van der Wel}, A., {Kelson}, D.~D., {Franx}, M., \&
  {Illingworth}, G.~D. 2010, \apj, 724, 714

\bibitem[{{Hopkins} {et~al.}(2010){Hopkins}, {Bundy}, {Hernquist}, {Wuyts}, \&
  {Cox}}]{Hopkins2010}
{Hopkins}, P.~F., {Bundy}, K., {Hernquist}, L., {Wuyts}, S., \& {Cox}, T.~J.
  2010, \mnras, 401, 1099

\bibitem[{{Hyde} \& {Bernardi}(2009)}]{Hyde2009b}
{Hyde}, J.~B., \& {Bernardi}, M. 2009, \mnras, 396, 1171

\bibitem[{{Ilbert} {et~al.}(2006){Ilbert}, {Arnouts}, {McCracken},
  {Bolzonella}, {Bertin}, {Le F{\`e}vre}, {Mellier}, {Zamorani}, {Pell{\`o}},
  {Iovino}, {Tresse}, {Le Brun}, {Bottini}, {Garilli}, {Maccagni}, {Picat},
  {Scaramella}, {Scodeggio}, {Vettolani}, {Zanichelli}, {Adami}, {Bardelli},
  {Cappi}, {Charlot}, {Ciliegi}, {Contini}, {Cucciati}, {Foucaud}, {Franzetti},
  {Gavignaud}, {Guzzo}, {Marano}, {Marinoni}, {Mazure}, {Meneux}, {Merighi},
  {Paltani}, {Pollo}, {Pozzetti}, {Radovich}, {Zucca}, {Bondi}, {Bongiorno},
  {Busarello}, {de La Torre}, {Gregorini}, {Lamareille}, {Mathez}, {Merluzzi},
  {Ripepi}, {Rizzo}, \& {Vergani}}]{Ilbert2006b}
{Ilbert}, O., {et~al.} 2006, \aap, 457, 841

\bibitem[{{Kauffmann} {et~al.}(2003){Kauffmann}, {Heckman}, {White}, {Charlot},
  {Tremonti}, {Brinchmann}, {Bruzual}, {Peng}, {Seibert}, {Bernardi},
  {Blanton}, {Brinkmann}, {Castander}, {Cs{\'a}bai}, {Fukugita}, {Ivezic},
  {Munn}, {Nichol}, {Padmanabhan}, {Thakar}, {Weinberg}, \&
  {York}}]{Kauffmann2003a}
{Kauffmann}, G., {et~al.} 2003, \mnras, 341, 33

\bibitem[{{Khochfar} \& {Silk}(2006)}]{Khochfar2006}
{Khochfar}, S., \& {Silk}, J. 2006, \apjl, 648, L21

\bibitem[{{Koekemoer} {et~al.}(2007){Koekemoer}, {Aussel}, {Calzetti}, {Capak},
  {Giavalisco}, {Kneib}, {Leauthaud}, {Le F{\`e}vre}, {McCracken}, {Massey},
  {Mobasher}, {Rhodes}, {Scoville}, \& {Shopbell}}]{Koekemoer2007}
{Koekemoer}, A.~M., {et~al.} 2007, \apjs, 172, 196

\bibitem[{{Koleva} {et~al.}(2009){Koleva}, {Prugniel}, {Bouchard}, \&
  {Wu}}]{Koleva2009}
{Koleva}, M., {Prugniel}, P., {Bouchard}, A., \& {Wu}, Y. 2009, \aap, 501, 1269

\bibitem[{{Le Borgne} {et~al.}(2004){Le Borgne}, {Rocca-Volmerange},
  {Prugniel}, {Lan{\c c}on}, {Fioc}, \& {Soubiran}}]{LeBorgne2004}
{Le Borgne}, D., {Rocca-Volmerange}, B., {Prugniel}, P., {Lan{\c c}on}, A.,
  {Fioc}, M., \& {Soubiran}, C. 2004, \aap, 425, 881

\bibitem[{{Maraston} \& {Str{\"o}mb{\"a}ck}(2011)}]{Maraston2011}
{Maraston}, C., \& {Str{\"o}mb{\"a}ck}, G. 2011, \mnras, 418, 2785

\bibitem[{{Misgeld} \& {Hilker}(2011)}]{Misgeld2011}
{Misgeld}, I., \& {Hilker}, M. 2011, \mnras, 414, 3699

\bibitem[{{Muzzin} {et~al.}(2013){Muzzin}, {Marchesini}, {Stefanon}, {Franx},
  {Milvang-Jensen}, {Dunlop}, {Fynbo}, {Brammer}, {Labb{\'e}}, \& {van
  Dokkum}}]{Muzzin2013a}
{Muzzin}, A., {et~al.} 2013, \apjs, 206, 8

\bibitem[{{Naab} {et~al.}(2009){Naab}, {Johansson}, \& {Ostriker}}]{Naab2009}
{Naab}, T., {Johansson}, P.~H., \& {Ostriker}, J.~P. 2009, \apjl, 699, L178

\bibitem[{{Newman} {et~al.}(2012){Newman}, {Genzel}, {F{\"o}rster-Schreiber},
  {Shapiro Griffin}, {Mancini}, {Lilly}, {Renzini}, {Bouch{\'e}}, {Burkert},
  {Buschkamp}, {Carollo}, {Cresci}, {Davies}, {Eisenhauer}, {Genel}, {Hicks},
  {Kurk}, {Lutz}, {Naab}, {Peng}, {Sternberg}, {Tacconi}, {Vergani}, {Wuyts},
  \& {Zamorani}}]{Newman2012}
{Newman}, S.~F., {et~al.} 2012, \apj, 761, 43

\bibitem[{{Padmanabhan} {et~al.}(2008){Padmanabhan}, {Schlegel}, {Finkbeiner},
  {Barentine}, {Blanton}, {Brewington}, {Gunn}, {Harvanek}, {Hogg},
  {Ivezi{\'c}}, {Johnston}, {Kent}, {Kleinman}, {Knapp}, {Krzesinski}, {Long},
  {Neilsen}, {Nitta}, {Loomis}, {Lupton}, {Roweis}, {Snedden}, {Strauss}, \&
  {Tucker}}]{Padmanabhan2008}
{Padmanabhan}, N., {et~al.} 2008, \apj, 674, 1217

\bibitem[{{Poggianti} {et~al.}(2013){Poggianti}, {Moretti}, {Calvi},
  {D'Onofrio}, {Valentinuzzi}, {Fritz}, \& {Renzini}}]{Poggianti2013}
{Poggianti}, B.~M., {Moretti}, A., {Calvi}, R., {D'Onofrio}, M.,
  {Valentinuzzi}, T., {Fritz}, J., \& {Renzini}, A. 2013, \apj, 777, 125

\bibitem[{{Saglia} {et~al.}(2010){Saglia}, {S{\'a}nchez-Bl{\'a}zquez},
  {Bender}, {Simard}, {Desai}, {Arag{\'o}n-Salamanca}, {Milvang-Jensen},
  {Halliday}, {Jablonka}, {Noll}, {Poggianti}, {Clowe}, {De Lucia},
  {Pell{\'o}}, {Rudnick}, {Valentinuzzi}, {White}, \& {Zaritsky}}]{Saglia2010}
{Saglia}, R.~P., {et~al.} 2010, \aap, 524, A6

\bibitem[{{S{\'a}nchez-Bl{\'a}zquez} {et~al.}(2006){S{\'a}nchez-Bl{\'a}zquez},
  {Peletier}, {Jim{\'e}nez-Vicente}, {Cardiel}, {Cenarro},
  {Falc{\'o}n-Barroso}, {Gorgas}, {Selam}, \&
  {Vazdekis}}]{Sanchez-Blazquez2006}
{S{\'a}nchez-Bl{\'a}zquez}, P., {et~al.} 2006, \mnras, 371, 703

\bibitem[{{Sargent} {et~al.}(2007){Sargent}, {Carollo}, {Lilly}, {Scarlata},
  {Feldmann}, {Kampczyk}, {Koekemoer}, {Scoville}, {Kneib}, {Leauthaud},
  {Massey}, {Rhodes}, {Tasca}, {Capak}, {McCracken}, {Porciani}, {Renzini},
  {Taniguchi}, {Thompson}, \& {Sheth}}]{Sargent2007}
{Sargent}, M.~T., {et~al.} 2007, \apjs, 172, 434

\bibitem[{{Sarzi} {et~al.}(2006){Sarzi}, {Falc{\'o}n-Barroso}, {Davies},
  {Bacon}, {Bureau}, {Cappellari}, {de Zeeuw}, {Emsellem}, {Fathi},
  {Krajnovi{\'c}}, {Kuntschner}, {McDermid}, \& {Peletier}}]{Sarzi2006}
{Sarzi}, M., {et~al.} 2006, \mnras, 366, 1151

\bibitem[{{Saulder} {et~al.}(2013){Saulder}, {Mieske}, {Zeilinger}, \&
  {Chilingarian}}]{Saulder2013}
{Saulder}, C., {Mieske}, S., {Zeilinger}, W.~W., \& {Chilingarian}, I. 2013,
  \aap, 557, A21

\bibitem[{{Saulder} {et~al.}(2015){Saulder}, {van den Bosch}, \&
  {Mieske}}]{Saulder2015}
{Saulder}, C., {van den Bosch}, R.~C.~E., \& {Mieske}, S. 2015, ArXiv e-prints

\bibitem[{{Schechter}(2015)}]{Schechter2015}
{Schechter}, P.~L. 2015, ArXiv e-prints

\bibitem[{{Scoville} {et~al.}(2007){Scoville}, {Aussel}, {Brusa}, {Capak},
  {Carollo}, {Elvis}, {Giavalisco}, {Guzzo}, {Hasinger}, {Impey}, {Kneib},
  {LeFevre}, {Lilly}, {Mobasher}, {Renzini}, {Rich}, {Sanders}, {Schinnerer},
  {Schminovich}, {Shopbell}, {Taniguchi}, \& {Tyson}}]{Scoville2007}
{Scoville}, N., {et~al.} 2007, \apjs, 172, 1

\bibitem[{{Sersic}(1968)}]{Sersic1968}
{Sersic}, J.~L. 1968, {Atlas de galaxias australes} (Observatorio Astronomico,
  Universidad Nacional de Cordoba)

\bibitem[{Shlens(2014)}]{Shlens2014}
Shlens, J. 2014, arXiv preprint arXiv:1404.1100

\bibitem[{{Simard} {et~al.}(2002){Simard}, {Willmer}, {Vogt}, {Sarajedini},
  {Phillips}, {Weiner}, {Koo}, {Im}, {Illingworth}, \& {Faber}}]{Simard2002}
{Simard}, L., {et~al.} 2002, \apjs, 142, 1

\bibitem[{{Smee} {et~al.}(2013){Smee}, {Gunn}, {Uomoto}, {Roe}, {Schlegel},
  {Rockosi}, {Carr}, {Leger}, {Dawson}, {Olmstead}, {Brinkmann}, {Owen},
  {Barkhouser}, {Honscheid}, {Harding}, {Long}, {Lupton}, {Loomis}, {Anderson},
  {Annis}, {Bernardi}, {Bhardwaj}, {Bizyaev}, {Bolton}, {Brewington}, {Briggs},
  {Burles}, {Burns}, {Castander}, {Connolly}, {Davenport}, {Ebelke}, {Epps},
  {Feldman}, {Friedman}, {Frieman}, {Heckman}, {Hull}, {Knapp}, {Lawrence},
  {Loveday}, {Mannery}, {Malanushenko}, {Malanushenko}, {Merrelli}, {Muna},
  {Newman}, {Nichol}, {Oravetz}, {Pan}, {Pope}, {Ricketts}, {Shelden},
  {Sandford}, {Siegmund}, {Simmons}, {Smith}, {Snedden}, {Schneider},
  {SubbaRao}, {Tremonti}, {Waddell}, \& {York}}]{Smee2013}
{Smee}, S.~A., {et~al.} 2013, \aj, 146, 32

\bibitem[{{Stoughton} {et~al.}(2002){Stoughton}, {Lupton}, {Bernardi},
  {Blanton}, {Burles}, {Castander}, {Connolly}, {Eisenstein}, {Frieman},
  {Hennessy}, {Hindsley}, {Ivezi{\'c}}, {Kent}, {Kunszt}, {Lee}, {Meiksin},
  {Munn}, {Newberg}, {Nichol}, {Nicinski}, {Pier}, {Richards}, {Richmond},
  {Schlegel}, {Smith}, {Strauss}, {SubbaRao}, {Szalay}, {Thakar}, {Tucker},
  {Vanden Berk}, {Yanny}, {Adelman}, {Anderson}, {Anderson}, {Annis},
  {Bahcall}, {Bakken}, {Bartelmann}, {Bastian}, {Bauer}, {Berman},
  {B{\"o}hringer}, {Boroski}, {Bracker}, {Briegel}, {Briggs}, {Brinkmann},
  {Brunner}, {Carey}, {Carr}, {Chen}, {Christian}, {Colestock}, {Crocker},
  {Csabai}, {Czarapata}, {Dalcanton}, {Davidsen}, {Davis}, {Dehnen},
  {Dodelson}, {Doi}, {Dombeck}, {Donahue}, {Ellman}, {Elms}, {Evans}, {Eyer},
  {Fan}, {Federwitz}, {Friedman}, {Fukugita}, {Gal}, {Gillespie}, {Glazebrook},
  {Gray}, {Grebel}, {Greenawalt}, {Greene}, {Gunn}, {de Haas}, {Haiman},
  {Haldeman}, {Hall}, {Hamabe}, {Hansen}, {Harris}, {Harris}, {Harvanek},
  {Hawley}, {Hayes}, {Heckman}, {Helmi}, {Henden}, {Hogan}, {Hogg}, {Holmgren},
  {Holtzman}, {Huang}, {Hull}, {Ichikawa}, {Ichikawa}, {Johnston}, {Kauffmann},
  {Kim}, {Kimball}, {Kinney}, {Klaene}, {Kleinman}, {Klypin}, {Knapp},
  {Korienek}, {Krolik}, {Kron}, {Krzesi{\'n}ski}, {Lamb}, {Leger},
  {Limmongkol}, {Lindenmeyer}, {Long}, {Loomis}, {Loveday}, {MacKinnon},
  {Mannery}, {Mantsch}, {Margon}, {McGehee}, {McKay}, {McLean}, {Menou},
  {Merelli}, {Mo}, {Monet}, {Nakamura}, {Narayanan}, {Nash}, {Neilsen},
  {Newman}, {Nitta}, {Odenkirchen}, {Okada}, {Okamura}, {Ostriker}, {Owen},
  {Pauls}, {Peoples}, {Peterson}, {Petravick}, {Pope}, {Pordes}, {Postman},
  {Prosapio}, {Quinn}, {Rechenmacher}, {Rivetta}, {Rix}, {Rockosi}, {Rosner},
  {Ruthmansdorfer}, {Sandford}, {Schneider}, {Scranton}, {Sekiguchi}, {Sergey},
  {Sheth}, {Shimasaku}, {Smee}, {Snedden}, {Stebbins}, {Stubbs}, {Szapudi},
  {Szkody}, {Szokoly}, {Tabachnik}, {Tsvetanov}, {Uomoto}, {Vogeley}, {Voges},
  {Waddell}, {Walterbos}, {Wang}, {Watanabe}, {Weinberg}, {White}, {White},
  {Wilhite}, {Wolfe}, {Yasuda}, {York}, {Zehavi}, \& {Zheng}}]{Stoughton2002}
{Stoughton}, C., {et~al.} 2002, \aj, 123, 485

\bibitem[{{Taylor} {et~al.}(2010){Taylor}, {Franx}, {Glazebrook}, {Brinchmann},
  {van der Wel}, \& {van Dokkum}}]{Taylor2010}
{Taylor}, E.~N., {Franx}, M., {Glazebrook}, K., {Brinchmann}, J., {van der
  Wel}, A., \& {van Dokkum}, P.~G. 2010, \apj, 720, 723

\bibitem[{{Thomas} {et~al.}(2013){Thomas}, {Steele}, {Maraston}, {Johansson},
  {Beifiori}, {Pforr}, {Str{\"o}mb{\"a}ck}, {Tremonti}, {Wake}, {Bizyaev},
  {Bolton}, {Brewington}, {Brownstein}, {Comparat}, {Kneib}, {Malanushenko},
  {Malanushenko}, {Oravetz}, {Pan}, {Parejko}, {Schneider}, {Shelden},
  {Simmons}, {Snedden}, {Tanaka}, {Weaver}, \& {Yan}}]{Thomas2013}
{Thomas}, D., {et~al.} 2013, \mnras, 431, 1383

\bibitem[{{Toft} {et~al.}(2007){Toft}, {van Dokkum}, {Franx}, {Labbe},
  {F{\"o}rster Schreiber}, {Wuyts}, {Webb}, {Rudnick}, {Zirm}, {Kriek}, {van
  der Werf}, {Blakeslee}, {Illingworth}, {Rix}, {Papovich}, \&
  {Moorwood}}]{Toft2007}
{Toft}, S., {et~al.} 2007, \apj, 671, 285

\bibitem[{{Treu} {et~al.}(2005){Treu}, {Ellis}, {Liao}, \& {van
  Dokkum}}]{Treu2005}
{Treu}, T., {Ellis}, R.~S., {Liao}, T.~X., \& {van Dokkum}, P.~G. 2005, \apjl,
  622, L5

\bibitem[{{Trujillo} {et~al.}(2009){Trujillo}, {Cenarro}, {de
  Lorenzo-C{\'a}ceres}, {Vazdekis}, {de la Rosa}, \& {Cava}}]{Trujillo2009}
{Trujillo}, I., {Cenarro}, A.~J., {de Lorenzo-C{\'a}ceres}, A., {Vazdekis}, A.,
  {de la Rosa}, I.~G., \& {Cava}, A. 2009, \apjl, 692, L118

\bibitem[{{Trujillo} {et~al.}(2007){Trujillo}, {Conselice}, {Bundy}, {Cooper},
  {Eisenhardt}, \& {Ellis}}]{Trujillo2007}
{Trujillo}, I., {Conselice}, C.~J., {Bundy}, K., {Cooper}, M.~C., {Eisenhardt},
  P., \& {Ellis}, R.~S. 2007, \mnras, 382, 109

\bibitem[{{Valentinuzzi} {et~al.}(2010){Valentinuzzi}, {Fritz}, {Poggianti},
  {Cava}, {Bettoni}, {Fasano}, {D'Onofrio}, {Couch}, {Dressler}, {Moles},
  {Moretti}, {Omizzolo}, {Kj{\ae}rgaard}, {Vanzella}, \&
  {Varela}}]{Valentinuzzi2010}
{Valentinuzzi}, T., {et~al.} 2010, \apj, 712, 226

\bibitem[{{van de Sande} {et~al.}(2014){van de Sande}, {Kriek}, {Franx},
  {Bezanson}, \& {van Dokkum}}]{vandeSande2014}
{van de Sande}, J., {Kriek}, M., {Franx}, M., {Bezanson}, R., \& {van Dokkum},
  P.~G. 2014, \apjl, 793, L31

\bibitem[{{van der Wel} {et~al.}(2014){van der Wel}, {Franx}, {van Dokkum},
  {Skelton}, {Momcheva}, {Whitaker}, {Brammer}, {Bell}, {Rix}, {Wuyts},
  {Ferguson}, {Holden}, {Barro}, {Koekemoer}, {Chang}, {McGrath},
  {H{\"a}ussler}, {Dekel}, {Behroozi}, {Fumagalli}, {Leja}, {Lundgren},
  {Maseda}, {Nelson}, {Wake}, {Patel}, {Labb{\'e}}, {Faber}, {Grogin}, \&
  {Kocevski}}]{vanderWel2014}
{van der Wel}, A., {et~al.} 2014, \apj, 788, 28

\bibitem[{{van Dokkum} {et~al.}(2008){van Dokkum}, {Franx}, {Kriek}, {Holden},
  {Illingworth}, {Magee}, {Bouwens}, {Marchesini}, {Quadri}, {Rudnick},
  {Taylor}, \& {Toft}}]{vanDokkum2008b}
{van Dokkum}, P.~G., {et~al.} 2008, \apjl, 677, L5

\bibitem[{{van Dokkum} \& {van der Marel}(2007)}]{vanDokkum2007}
{van Dokkum}, P.~G., \& {van der Marel}, R.~P. 2007, \apj, 655, 30

\bibitem[{{Wake} {et~al.}(2012){Wake}, {van Dokkum}, \& {Franx}}]{Wake2012}
{Wake}, D.~A., {van Dokkum}, P.~G., \& {Franx}, M. 2012, \apjl, 751, L44

\bibitem[{{Wellons} {et~al.}(2015){Wellons}, {Torrey}, {Ma}, {Rodriguez-Gomez},
  {Vogelsberger}, {Kriek}, {van Dokkum}, {Nelson}, {Genel}, {Pillepich},
  {Springel}, {Sijacki}, {Snyder}, {Nelson}, {Sales}, \&
  {Hernquist}}]{Wellons2015a}
{Wellons}, S., {et~al.} 2015, \mnras, 449, 361

\bibitem[{{Woods} {et~al.}(2010){Woods}, {Geller}, {Kurtz}, {Westra},
  {Fabricant}, \& {Dell'Antonio}}]{Woods2010}
{Woods}, D.~F., {Geller}, M.~J., {Kurtz}, M.~J., {Westra}, E., {Fabricant},
  D.~G., \& {Dell'Antonio}, I. 2010, \aj, 139, 1857

\bibitem[{{York} {et~al.}(2000){York}, {Adelman}, {Anderson}, {Anderson},
  {Annis}, {Bahcall}, {Bakken}, {Barkhouser}, {Bastian}, {Berman}, {Boroski},
  {Bracker}, {Briegel}, {Briggs}, {Brinkmann}, {Brunner}, {Burles}, {Carey},
  {Carr}, {Castander}, {Chen}, {Colestock}, {Connolly}, {Crocker}, {Csabai},
  {Czarapata}, {Davis}, {Doi}, {Dombeck}, {Eisenstein}, {Ellman}, {Elms},
  {Evans}, {Fan}, {Federwitz}, {Fiscelli}, {Friedman}, {Frieman}, {Fukugita},
  {Gillespie}, {Gunn}, {Gurbani}, {de Haas}, {Haldeman}, {Harris}, {Hayes},
  {Heckman}, {Hennessy}, {Hindsley}, {Holm}, {Holmgren}, {Huang}, {Hull},
  {Husby}, {Ichikawa}, {Ichikawa}, {Ivezi{\'c}}, {Kent}, {Kim}, {Kinney},
  {Klaene}, {Kleinman}, {Kleinman}, {Knapp}, {Korienek}, {Kron}, {Kunszt},
  {Lamb}, {Lee}, {Leger}, {Limmongkol}, {Lindenmeyer}, {Long}, {Loomis},
  {Loveday}, {Lucinio}, {Lupton}, {MacKinnon}, {Mannery}, {Mantsch}, {Margon},
  {McGehee}, {McKay}, {Meiksin}, {Merelli}, {Monet}, {Munn}, {Narayanan},
  {Nash}, {Neilsen}, {Neswold}, {Newberg}, {Nichol}, {Nicinski}, {Nonino},
  {Okada}, {Okamura}, {Ostriker}, {Owen}, {Pauls}, {Peoples}, {Peterson},
  {Petravick}, {Pier}, {Pope}, {Pordes}, {Prosapio}, {Rechenmacher}, {Quinn},
  {Richards}, {Richmond}, {Rivetta}, {Rockosi}, {Ruthmansdorfer}, {Sandford},
  {Schlegel}, {Schneider}, {Sekiguchi}, {Sergey}, {Shimasaku}, {Siegmund},
  {Smee}, {Smith}, {Snedden}, {Stone}, {Stoughton}, {Strauss}, {Stubbs},
  {SubbaRao}, {Szalay}, {Szapudi}, {Szokoly}, {Thakar}, {Tremonti}, {Tucker},
  {Uomoto}, {Vanden Berk}, {Vogeley}, {Waddell}, {Wang}, {Watanabe},
  {Weinberg}, {Yanny}, \& {Yasuda}}]{York2000}
{York}, D.~G., {et~al.} 2000, \aj, 120, 1579

\bibitem[{{Zahid} {et~al.}(2015){Zahid}, {Damjanov}, {Geller}, \&
  {Chilingarian}}]{Zahid2015a}
{Zahid}, H.~J., {Damjanov}, I., {Geller}, M.~J., \& {Chilingarian}, I. 2015,
  \apj, 806, 122

\bibitem[{{Zirm} {et~al.}(2007){Zirm}, {van der Wel}, {Franx}, {Labb{\'e}},
  {Trujillo}, {van Dokkum}, {Toft}, {Daddi}, {Rudnick}, {Rix},
  {R{\"o}ttgering}, \& {van der Werf}}]{Zirm2007}
{Zirm}, A.~W., {et~al.} 2007, \apj, 656, 66

\end{thebibliography}

\LongTables
\begin{deluxetable*}{ccccccc}
\tablewidth{350pt}
\tablecaption{Fundamental Plane Properties}
\tablehead{\colhead{RA} &\colhead{Dec} &\colhead{Redshift} & \colhead{$R_e$} &\colhead{log$(M_\ast/M_\odot)$ } & \colhead{$M_r$} & \colhead{$\sigma$} \\
                           \colhead{}  & \colhead{}  &\colhead{}  &       \colhead{[kpc]} &               \colhead{}                &  \colhead{[mags]} & \colhead{km s$^{-1}$ }    }
\startdata
150.03556 & 2.24671 & 0.2827 & 4.19 & 11.12 & -22.12 & 193 $\pm$ 22 \\
\enddata
\label{tab:prop}
\tablecomments{A machine readable table is available upon request from HJZ. The corrections described in Section 2 are applied to the data.}
\end{deluxetable*}

 \end{document}